\documentclass[journal=jpcafh,manuscript=article]{achemso}

\usepackage[version=3]{mhchem} % Formula subscripts using \ce{}
\usepackage{multirow}
\usepackage{xr}
\author{Vladislav Slama}
\affiliation[EPFL]{Laboratory of Computational Chemistry and Biochemistry, École Polytechnique Fédérale de Lausanne,  1015 Lausanne, Switzerland}  %, Lausanne, Switzerland

\author{Camila Negrete-Vergara}
\affiliation[Bern]{Department of Chemistry, Biochemistry and Pharmaceutical Sciences, W. Inaebnit Laboratory for Molecular Quantum Materials and WSS-Research Center for Molecular Quantum Systems, University of Bern, Freiestrasse 3, 3012 Bern, Switzerland}  %, Bern, Switzerland

\author{Elnaz Zyaee}
\affiliation[inst3]
{Institute of Applied Physics, University of Bern, Sidlerstrasse 5, 3012 Bern, Switzerland} %, Bern, Switzerland

\author{Silvio Decurtins}
\affiliation[Bern]{Department of Chemistry, Biochemistry and Pharmaceutical Sciences, W. Inaebnit Laboratory for Molecular Quantum Materials and WSS-Research Center for Molecular Quantum Systems, University of Bern, Freiestrasse 3, 3012 Bern, Switzerland}  %, Bern, Switzerland

\author{Pascal Manuel H\"{a}nzi}
\affiliation[inst3]
{Institute of Applied Physics, University of Bern, Sidlerstrasse 5, 3012 Bern, Switzerland} %, Bern, Switzerland

\author{Thomas Feurer}
\affiliation[inst3]
{Institute of Applied Physics, University of Bern, Sidlerstrasse 5, 3012 Bern, Switzerland} %, Bern, Switzerland

\author{Shi-Xia Liu}
\email{shi-xia.liu@unibe.ch}
\affiliation[Bern]{Department of Chemistry, Biochemistry and Pharmaceutical Sciences, W. Inaebnit Laboratory for Molecular Quantum Materials and WSS-Research Center for Molecular 
Quantum Systems, University of Bern, Freiestrasse 3, 3012 Bern, Switzerland} %, Bern, Switzerland

\author{Ursula Rothlisberger}
\email{ursula.roethlisberger@epfl.ch}
\affiliation[EPFL]{Laboratory of Computational Chemistry and Biochemistry, École Polytechnique Fédérale de Lausanne,  1015 Lausanne, Switzerland} % , Lausanne, Switzerland

\title[An \textsf{achemso} demo]
  {An accurate theoretical framework for the optical and electronic properties of paracyclophanes}

\abbreviations{IR,NMR,UV}

\begin{document}

\begin{tocentry}

\begin{figure} [H] %[ht!]
    \centering
    \includegraphics[width=1.0\columnwidth]{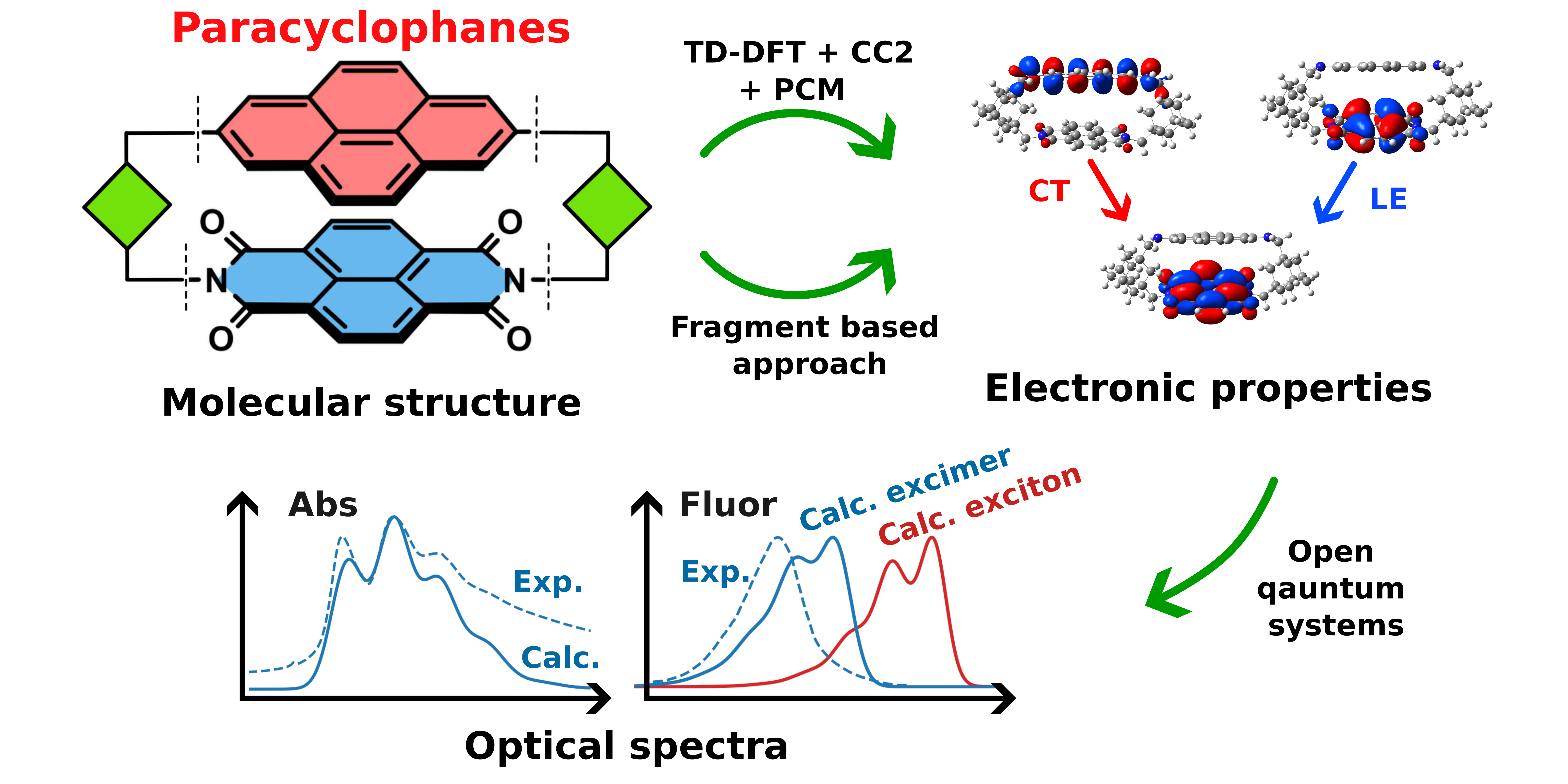}
\end{figure}

\end{tocentry}

\begin{abstract}

Aromatic $\pi$–stacking interactions play an important role in both natural and artificial systems, influencing processes such as charge separation in photosynthesis and charge transport in organic semiconductors. Controlling the geometry and distance between aromatic units is therefore crucial for tuning intermolecular interactions and charge-transfer efficiency. Due to their well-defined stacking geometry, paracyclophanes (PCPs) composed of two or more aromatic units connected by rigid linkers, provide an ideal platform for a systematic study of such effects. Despite extensive experimental studies of PCPs, a comprehensive and quantitatively validated theoretical description linking the structure with the electronic and optical properties is still missing. Here, we present an extensive computational and experimental investigation of the electronic and optical properties of homo-PCPs containing naphthalene diimide (NDI) or pyrene chromophores linked by bridges of varying length and rigidity. We introduce a robust methodology for an accurate simulation of the absorption and fluorescence spectra of PCPs based on a combined TD-DFT and CC2 approach, achieving excellent quantitative agreement with experiment. We also present and validate a fragment-based description of PCPs using the Frenkel exciton model. Such approach is valuable not only for interpretation of the electronic and optical properties of PCPs, but it can also significantly reduce the cost of the calculation while maintaining the accuracy of the supermolecular approach. This work establishes a quantitatively reliable framework linking structure, excitonic coupling, and charge-transfer interactions in PCPs with optical properties, providing design principles for next-generation optoelectronic materials. 

\end{abstract}

\section{Introduction}

Aromatic stacking interactions play an important role in many fields ranging from biology to supramolecular chemistry and materials science. In biological systems $\pi$-stacking contributes to protein stability, ligand-receptor binding (e.g. in dopamine receptors) \cite{Dougherty2013,Bueschbell2019},and and molecular orbital overlap of aromatic systems is crucial for charge separation in the photosynthetic reaction center \cite{ALLEN19985}. In artificial systems, such as organic electronic devices or artificial light harvesting systems, $\pi$-stacking enhances molecular orbital overlap, which is critical for charge transport and charge-transfer (CT) state formation \cite{Wasielewski2009,OSAKA2015,Facchetti2011,Khasbaatar2023}. The efficiency of CT state formation and intermolecular charge transport in organic semiconductors is governed by electronic coupling \cite{Wang2012}, which depends on the degree of molecular orbital overlap between neighboring molecules. This overlap, in turn, is determined by the relative distance and orientation of the molecules.
%The extend of the molecular orbital overlap is determined by the mutual distance and orientation of the neighboring molecules. 
Therefore, significant efforts have been devoted towards the development of molecular systems that exhibit close packing and well-defined orientations to tailor intermolecular interaction to specific applications.

A promising class of compounds featuring face-to-face aromatic stacking and well-defined mutual geometry are paracyclophanes (PCPs) \cite{Wu2019,Hopf2008,Lyssenko2003}. These molecules consist of two or more aromatic units connected by linkers (e.g. alkyl chains or phenylene bridges), resulting in a rigid structure. The electronic and optical properties of PCPs are governed by both the choice of the aromatic subunits and the linkers, which defines the interunit distance as well as the relative orientation of the two aromatic systems. Homoparacyclophanes (homo-PCPs), composed of identical aromatic units, have been explored as building blocks for the construction of conductive polymers, molecular wires, and light-emitting diodes \cite{Hopf2008,Reznikova2021,GRYKIEN2016}. In contrast, heteroparacyclophanes formed by aromatic units with electron donor and acceptor properties exhibit pronounced CT interactions \cite{Wang2025}. The ability to tune CT state properties in these systems is particularly relevant for applications in organic photovoltaics (OPV) and organic field-effect transistors (OFET) \cite{Reznikova2021,Lin2018}.

Despite extensive experimental studies of PCPs, a comprehensive and quantitatively validated theoretical description linking molecular structure with electronic and optical properties, particularly for larger systems, is still missing. Most of the theoretical investigations to date have focused on ground-state properties and molecular orbital analyses, or time dependent density functional theory (TD-DFT) calculations of excited states in the gas phase \cite{Caramori2007,Wang2025,Bao2025,Lin2008}. The lack of quantitative accuracy is largely due to the challenges in the theoretical description of CT states, which are characterized by spatially separated electrons and holes with small molecular orbital overlap resulting in low transition dipole moments, and a strong dependence on geometry and environment. Conventional TD-DFT approaches employing generalized gradient approximation (GGA) or hybrid functionals often fail for CT states because of their incorrect asymptotic behavior, necessitating the use of higher-level methods with significantly greater computational cost \cite{Dreuw2004,Casida1998}. Although employing range-separated hybrid functionals partially mitigate this deficiency of the TD-DFT approach, in many cases they still do not achieve fully quantitative accuracy \cite{Slama2023,Begam2020}.

In this work, we present a computational methodology for a quantitative description of the electronic properties of PCPs. The redox properties are calculated using DFT-based approaches and validated against  experimental measurements. Excited state properties are evaluated using a combination of an approximate coupled cluster method (CC2) and linear-response TD-DFT with range-separated functionals. The calculated excited states are used to simulate absorption and fluorescence spectra and to rationalize the observed excited state dynamics of PCP systems. We also investigate solvent effects on both redox and excited-state properties and demonstrate that accurate inclusion of solvation is essential for a quantitative agreement with experimental data. Furthermore,  we show that the electronic properties of homo-PCPs can effectively be described as interacting monomeric units embedded in a polarizable environment, following the Frenkel exciton model. This approach substantially reduces computational cost, enables the treatment of larger systems, and facilitates the rational design of novel PCPs with tailored properties.

Finally, we validate the proposed methodology by comparing computed optical spectra with experimental results for a series of homo-PCPs containing naphthalene diimide (NDI) or pyrene units connected by linkers of varying length and rigidity, thereby covering a range of interunit distances and relative orientations. The resulting quantitative agreement confirms the robustness of the computational approach. Importantly, an accurate theoretical description of the electronic and optical properties of PCPs is not only essential for understanding their behavior in different environments but it is also critical for the rational design of new molecules with targeted functionalities. 

\section{Theoretical Methods}
To compare the electronic and optical properties of the monomeric NDI and pyrene reference compounds with those of the corresponding homo-PCPs, the molecular structures were optimized at the DFT level using the B3LYP functional and the 6-31+G(p,d) basis set. Solvent effects were included through the polarizable continuum model (PCM) with equilibrium solvation, employing dichloromethane (\ce{CH2Cl2}) as the solvent. 
Since the lowest excited state of pyrene with transition dipole moment oriented along its short molecular axis ($^{1}L_{b}$) exhibits partial double-excitation character and the pyrene with amine-containing linkers show partial charge-transfer (CT) character, a two-step procedure was employed to compute the excited-state properties of PCPs containing pyrene units. 

In the first step, excited state properties were calculated in the gas phase at CC2 level with a def2-TZVPD basis set as implemented in Turbomole 7.1.1 \cite{Turbo1989}, and at the TD-DFT level using the range-separated $\omega$B97XD functional with the same basis set as  implemented in Gaussian 16 \cite{g16}, both at the same DFT geometry. The CC2 method provides similar accuracy for the locally excited (LE) state energies for both the NDI and pyrene units (including states with partial double excitation character) as for the CT states in the PCPs. While TD-DFT with range separated hybrid functionals generally fails to reproduce the absolute energies of CT states and the excited states of pyrene  with partial double excitation character, it correctly describes their molecular orbital (MO) composition and symmetry. Using a natural transition orbital (NTO) analysis, CC2 excited states were mapped onto their TD-DFT counterparts. This mapping was used to obtain energy correction factors for the states computed at the TD-DFT level of theory. 

In the second step, excited states are recomputed at the TD-DFT level with the $\omega$B97XD functional including solvent effects at the same geometry as in the gas phase calculation, using the Gaussian 16 software package \cite{g16}. Solvent effects were incorporated within the PCM approach where the implicit non-equilibrium solvation was applied for the excited states, the linear response approach was used for LE states, and a state-specific solvation model was adopted for the states with significant CT character. Analysis of the MO composition of the excited states and the NTOs was used to map the solvated excited states to their gas phase counterparts. The final excitation energies of the solvated systems were obtained from the TD-DFT calculation with nonequlibrium solvation, corrected by the corresponding energy difference between TD-DFT and CC2 from the gas phase. This two-step approach effectively overcomes the inherent limitations of TD-DFT for CT states and yields quantitatively accurate excitation energies for solvated molecules without the need for empirical energy shifts. Conceptually, this method can be regarded as a CC2 gas-phase calculation corrected by solvent-induced energy shifts computed at the TD-DFT level. 

This two-step procedure was necessary because the current implementation of the CC2 method in Turbomole 7.1.1 \cite{Turbo1989} does not support explicit inclusion of solvent effects in a state-specific manner—a feature that turned out to be essential for an accurate description of solvent effects on CT states.

Optical spectra were simulated using the vertical gradient (VG) approximation combined with the cumulant expansion formalism \cite{Cignoni2022}. This method is widely applied in the modeling of organic molecules and super-molecular systems and accounts for vibronic effects in optical spectra without requiring extensive explicit sampling of the molecular geometries. It also enables combination of different theoretical methods for describing electronic excitations (CC2) and solvent effects (TD-DFT). Within this framework, the absorption $\alpha \left( \omega  \right) $ and fluorescence $ I \left( \omega  \right) $ spectra are obtained as a sum over excited states:

\begin{equation}
  \alpha \left( \omega  \right) \propto \omega \sum_{i} \left| \mu_{gi} \right|^{2} D_i\left(\omega \right),
  \label{eqn:Abs}
\end{equation}

\begin{equation}
  I \left( \omega  \right) \propto \omega^{2} \sum_{i} p_{i} \left| \mu_{gi} \right|^{2} \tilde{D}_i\left(\omega \right),
  \label{eqn:Fluor}
\end{equation}

where $p_i$ corresponds to the Boltzmann factor $p_i=\exp \left[-\frac{\varepsilon_i}{k_BT} \right]/\sum_{j}\exp \left[-\frac{\varepsilon_j}{k_BT} \right] $ and $\mu_{gi}$ is the transition dipole moment between the ground and $i$-th electronic excited state. 

The homogeneous lineshape $D_i\left(\omega \right)$  and fluorescence lineshape $\tilde{D}_i\left(\omega \right)$ are obtained in the cumulant expansion formalism as: 

\begin{equation}
  D_i\left(\omega \right) = \int^{\infty}_{-\infty} e^{-i\left( \omega_{gi} -\omega \right)t - g_i\left( t \right)} dt,
  \label{eqn:AbsLine}
\end{equation}

\begin{equation}
  \tilde{D}_i\left(\omega \right) = \int^{\infty}_{-\infty} e^{-i\left( \omega_{gi} -\omega \right)t +2i\lambda_i t - g^{\ast}_i\left( t \right)} dt,
  \label{eqn:FluorLine}
\end{equation}

where $g_i\left( t \right)$ is the lineshape function of the excited state $i$ and $\lambda_{i}$ is the corresponding reorganization energy. The lineshape function $g_i\left( t \right)$ can be obtained from the spectral density $J_i\left(\omega\right)$ of the excited state $i$ as:

\begin{equation}
  g_i\left( t \right) = \int^{\infty}_{0} d\omega \frac{J_i\left( \omega \right)}{\pi\omega^{2}} \left[ \coth\left( \frac{\beta\hbar\omega}{2} \right) \left( \cos\left(\omega t \right) - 1 \right) -i\left( \sin \left(\omega t \right) -\omega t \right) \right]
  \label{eqn:Lineshape}
\end{equation}

More details regarding the calculation of the spectral densities $J_i\left(\omega\right)$ are described in the Supporting Information (SI). 

\section{Experimental Methods}

NDI-Ada-NDI, NDI-tBuPh-NDI, and reference compound NDI-tBuPh were synthesized by procedures reported in the literature, with some modifications. The synthetic procedures and characterization for Pyrene-Ada and Pyrene-tBuPh are described in the (SI). UV-Vis absorption and fluorescence spectra of NDI-Ada-NDI, NDI-tBuPh-NDI, and reference compound NDI-tBuPh were recorded in \ce{CH2Cl2} at 10 $\mu$M concentration. UV-Vis absorption and fluorescence spectra of Pyrene-Ada and Pyrene-tBuPh were recorded in \ce{CH2Cl2} at 5 $\mu$M concentration. 

Cyclic voltametry (CV) was performed in a three-electrode cell with Pt working electrode, a glassy carbon counter-electrode, and a \ce{Ag}/\ce{AgCl} reference electrode. The electrochemical experiments were conducted under an oxygen-free atmosphere in dichloromethane (5 mL) with \ce{TBAPF_6} (0.1 M) as the supporting electrolyte and 1 mg of sample.

\subsection{Femtosecond Transient Absorption Spectroscopy (fTAS)}

Femtosecond transient absorption (TAS) measurements were performed using a Ti:sapphire regenerative amplifier system (800 nm central wavelength, 87 fs pulse duration, 4 kHz repetition rate, 0.55 mJ per pulse). Pump pulses were generated in a noncollinear optical parametric amplifier (NOPA) seeded with white light from the fundamental output. The NOPA output at 760 nm was frequency-doubled in a 0.1 mm BBO crystal to produce excitation pulses centered at 380 nm. The pump pulse energy at the sample was maintained at 0.5 µJ to avoid multiphoton effects. The probe pulse was generated as a supercontinuum (420–750 nm) in a 5 mm \ce{CaF2} plate and collinearly overlapped with the pump in the sample. The instrument response function (IRF) was determined to be 200 fs (FWHM). The temporal delay between pump and probe pulses was controlled with a motorized delay stage, covering delays up to 1.9 ns. Detailed information about the transient absorption measurements together with the schematic representation of the setup are provided in SI and Figure S4.

\section{Results and Discussion}

\subsection{Structural properties}

\begin{figure} [h] %[ht!]
    \centering
    \includegraphics[width=1.0\textwidth]{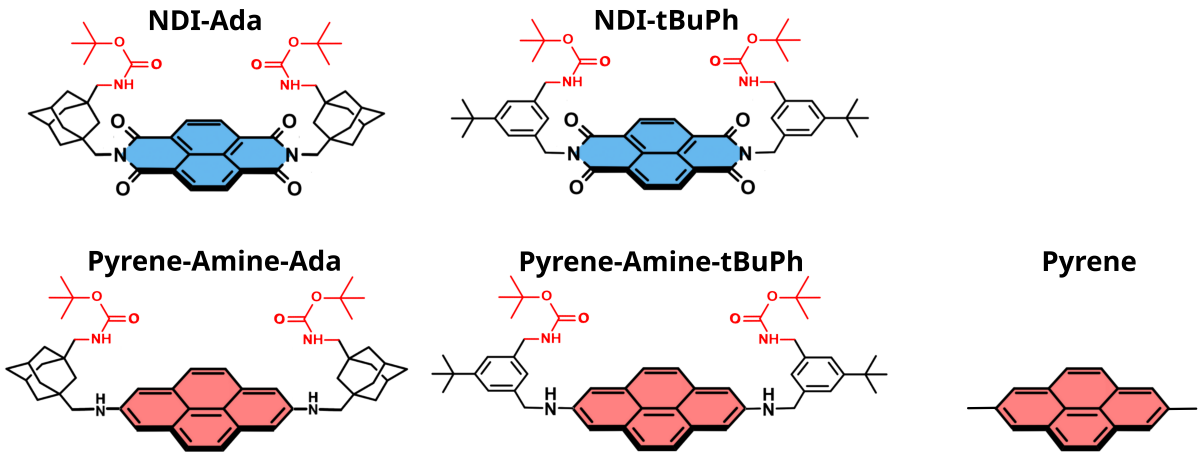}
    \caption{\label{fig:StructureMonomers} 
    \textbf{Chemical structure of the studied monomeric units.} NDI units are represented in blue, and pyrene ones in red. The red capping groups of the linkers are present in the synthesized molecules used for the experimental measurements, however, for the theoretical calculations the  groups highlighted in red are replaced by methyl groups. This does not influence optical or redox properties.
}
\end{figure}

The geometries of the monomeric reference compounds (Figure \ref{fig:StructureMonomers}) and the homo-PCPs were optimized using DFT, including solvent effects of \ce{CH2Cl2} via the PCM approach. The resulting PCP geometries are strongly influenced by both the structure of the individual aromatic units and the nature of the linkers. 

\begin{figure} [H] %[ht!]
    \centering
    \includegraphics[width=1.0\textwidth]{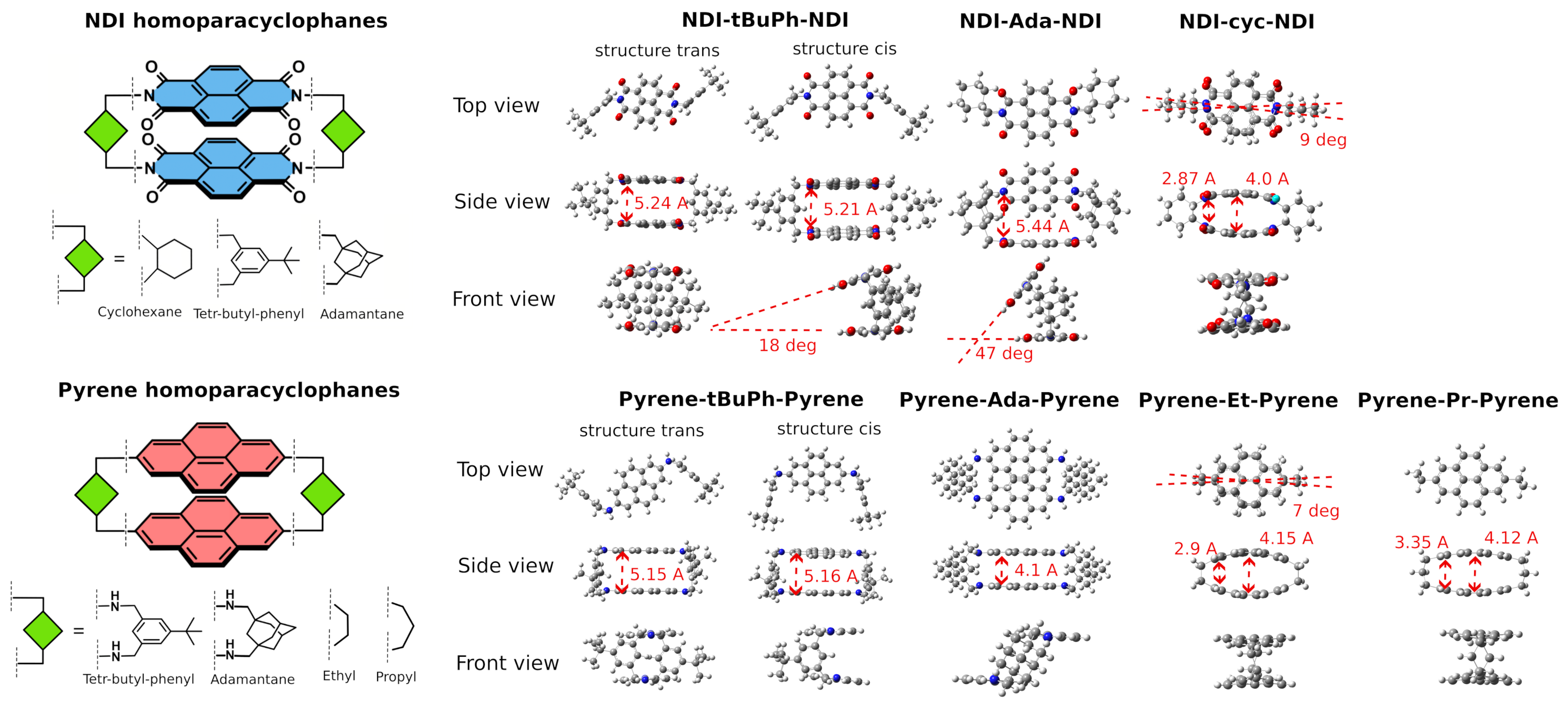}
    \caption{\label{fig:StructureParacyclophane} 
    \textbf{Structures of the studied homo-paracyclophanes.} Schematic representation of the paracyclophanes composed of two monomeric units and a linker (on the left) and the corresponding optimized  structures at DFT level in implicit DCM solvent.
}
\end{figure}

For the PCPs containing NDI units, three types of linkers were investigated: adamantane (Ada), \textit{tert}-butylphenyl (tBuPh) and cyclohexyl (cyc). The tBuPh linker provides higher flexibility for the mutual conformations of the aromatic units, allowing for the incorporation of units with slightly different lengths facilitating the parallel stacking of the NDI units. 
In contrast, the rigid Ada linker enforces a more constrained geometry, resulting in a pronounced inclination of the aromatic unit planes.
This difference mainly arises from the distinct volumes and shapes of the linkers. The bulky, nearly spherical Ada linker induces van der Waals repulsion with the adjacent aromatic units upon geometric fluctuations, thereby enforcing a more rigid structure. Conversely, the nearly planar tBuPh linker permits larger conformational fluctuations without steric clashes with the aromatic units.
For the NDI-tBuPh-NDI homodimer, two nearly isoenergetic lowest-energy conformers were identified, corresponding to the \textit{cis} and \textit{trans} orientations of the linkers, Figure \ref{fig:StructureParacyclophane}. The \textit{trans} orientation promotes nearly parallel stacking of the NDI units, while the \textit{cis} conformation leads to an inclination of approximatelly 18$^\circ$. The inclination is even more pronounced for the NDI-Ada-NDI homodimer, reaching a value around 47°. 

In the NDI homo-PCP with the shortest cyclohexane linker, the interunit distance near the linker is 2.9~Å, which is smaller than the sum of the van der Waals radii for two carbon atoms (approximately 3.4 Å) This close proximity results in steric repulsion between the NDI units, which suppresses inclination and induces bending of the overall structure. As a consequence, the interunit distance increases to approximately 4.0~Å at the center of the PCP, Figure \ref{fig:StructureParacyclophane}. This value is slightly larger than the interplanar distance observed in the crystalline state (around 3.3~Å), likely due to the absence of the intermolecular crystal packing constraints in solution. The calculated mutual rotation of 9$^\circ$ of the NDI units on the other hand agrees well with the previously reported crystal structure in Ref.~\citenum{Wu2014}.

A similar structural behavior is observed for PCPs containing pyrene units. The linkers examined in this case include ethyl, propyl, bis(aminomethyl)adamantane, and bis(aminomethyl)\textit{tert}-butylphenyl. The linkers bearing amine groups were selected because of their influence on the lowest excited state of pyrene; the amine acts as an electron donor, introducing partial CT character to this state. PCPs with  tBuPh and Ada linkers exhibit parallel arrangement of the pyrene units, with interplanar separations of 5.1 Å and 4.1 Å, respectively. As in the case of the NDI-based PCPs, the tBuPh-linked pyrene dimers also yield \textit{cis} and \textit{trans} conformers with similar energies. The shortest ethyl and propyl linkers lead to bent geometries, with interunit distances of about 4.1~Å at the center of the PCP and slightly shorter distances near the linkers—analogously to the behavior observed for the NDI–cyc–NDI system.

\subsection{Optical properties of the monomeric reference compounds}

\begin{figure} [H] %[ht!]
    \centering
    \includegraphics[width=0.87\columnwidth]{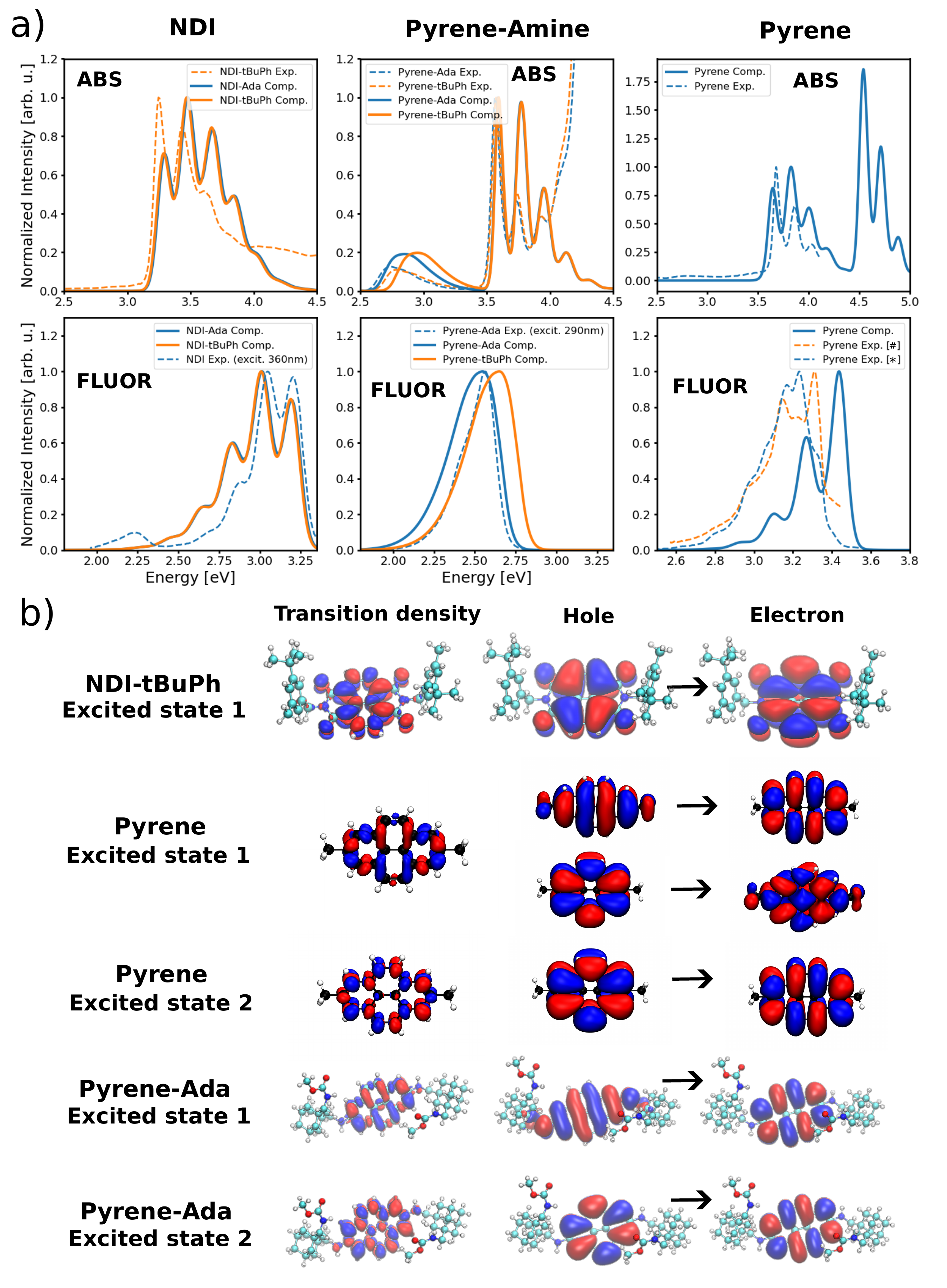}
    \caption{\label{fig:MonomerSpectra} 
    \textbf{Optical spectra and corresponding electronic excitations of the monomeric reference compounds.} \textbf{(a)} Computed and experimental absorption and fluorescence spectra of  NDI, pyrene and pyrene with tBuPh and Ada linkers. The computed spectra show excellent agreement with the experimental ones without inclusion of arbitrary shifts. The experimental spectrum for NDI-cyc-NDI was obtained from Ref.~\citenum{Wu2015}, and for Pyrene-Et-Pyrene from Ref.~\citenum{Staab1979}.  \textbf{(b)} Transition densities and natural transition orbitals for the lowest excited states of the monomeric reference compounds. The experimental spectra $^{\ast}$ and $^{\#}$ for pyrene were taken from Ref.~\citenum{Gac2025} and Ref.~\citenum{Galisinova2013}, respectively.
}
\end{figure}

Before analyzing the "double-decker" PCPs and elucidating the origin and dynamics of their excited states, it is instructive to first examine the excited-state properties of the individual molecular building blocks themselves. Studying the monomers separately allows for a clear distinction between inter-unit excitations within the PCP and intra-unit excitations localized on the individual monomeric units. The optical spectra of the NDI and pyrene monomers, together with a molecular orbital (MO) analysis of their lowest excited states, are presented in Figure~\ref{fig:MonomerSpectra}.

In the low-energy region (3.0–3.9~eV) of the NDI absorption spectrum, the dominant feature corresponds to the $\pi$–$\pi^{\ast}$ type $S_0 \rightarrow S_1$ with the transition dipole moment oriented along the longer symmetry axis, peaking near 3.25~eV. Natural transition orbital (NTO) analysis indicates that this transition is primarily composed (98$\%$) of the $HOMO \rightarrow LUMO$ single-electron excitation  (Figure \ref{fig:MonomerSpectra}). Both the hole and electron orbitals are localized on the NDI core, consistent with only a minimal spectral shift between NDI monomers with different linkers e.g. a tiny red shift of approximately 0.012~eV is observed for NDI with the tBuPh linker relative to the system with Ada linker. The fine structure in the optical spectrum arises from vibronic excitations of the C–C stretching modes of the $\pi$-conjugated system, with computed frequencies of 1446 and 1640~cm$^{-1}$ and corresponding reorganization energies of 1132 and 674~cm$^{-1}$, respectively. These values closely reproduce the experimentally observed vibronic peak separations of 1450 and $\sim$1580~cm$^{-1}$. The computed reorganization energies are slightly overestimated, leading to a lower intensity of the first vibronic peak in the simulated absorption spectrum compared to experiment, a well-known artifact of the vertical gradient approximation, which neglects small differences in normal modes between ground and excited states, and the DFT functional. The spectral shape can be further corrected by including Duschinsky rotation effects \cite{Duschinsky1937,Santoro2008,OlbrichKupka1983} (Figure~S1). Despite these minor deviations, the computed NDI absorption spectra is in very good quantitative agreement with the experimental spectra without the need for arbitrary shifts or rescaling of the excitation energies. Note that the deviation between calculated and measured spectra in the high-energy region is due to the fact that only the lowest excited states were included in the calculations.

Similar properties can be observed for the optical spectrum of pyrene. It is composed of three principal electronic transitions: the $^1L_b$ ($S_0 \rightarrow S_1$) state, composed of $HOMO-1 \rightarrow LUMO$ and $HOMO \rightarrow LUMO+1$ excitations; the $^1L_a$ state corresponding to ($S_0 \rightarrow S_2$) excitation, primarily a $HOMO \rightarrow LUMO$ transition; and the higher-lying $S_0 \rightarrow S_4$ transition, which gives rise to the intense UV absorption band around 4.5~eV. The first excited state $^1L_b$ possesses a weak transition dipole moment oriented along the short molecular axis, while the $^1L_a$ transition near 3.65~eV is polarized along the longer molecular axis. As in the case of NDI, TD-DFT slightly overestimates the reorganization energy, but the computed positions and relative intensities of both electronic and vibronic features are in very good agreement with experiment. Although the TD-DFT ($\omega$B97XD) calculations predict an incorrect ordering of the lowest excited states in solution, the correct ordering is restored upon applying the CC2-based energy corrections.
This discrepancy between CC2 and TD-DFT with range-separated hybrid functionals primarily originates from their different treatments of dynamic correlation and double excitations. In TD-DFT, correlation effects are incorporated in an approximate manner, and the adiabatic kernel used in linear-response TD-DFT cannot capture true two-electron dynamic correlation or the effects of double excitations \cite{Giesbertz2008,Elliott2011}. In contrast, the CC2 method includes double-excitation effects perturbatively and explicitly accounts for two-electron interactions, often yielding more accurate excitation energies when dynamic correlation with partial double-excitation character are significant \cite{Hellweg2008,Hattig2002}.
The simulated fluorescence spectra also reproduce the experimental shapes well, albeit blue-shifted by about 0.2~eV, primarily due to the overestimated excitation energy of the $^1L_b$ state, which experimentally lies near 3.3~eV \cite{Nakanishi1991}. A more accurate description of this state would require a higher-level method beyond the CC2 approach. 

For pyrene containing the amine-functionalized linkers (Ada and tBuPh), the low-energy region of the optical spectrum (2.6–3.5~eV) is dominated by a broad and featureless $S_0 \rightarrow S_1$ transition with a maximum near 2.74~eV. This transition corresponds to the $^1L_b$ excitation polarized along the shorter molecular axis, but with partial contribution from an intramolecular charge-transfer (ICT) component \cite{Crawford2011,Merz2017,Kurata2016,Jana2020,Li2014,Kurata2017}, as illustrated in Figure~\ref{fig:MonomerSpectra}b. NTO analysis reveals that the amine substituent acts as an electron donor: the hole orbital is delocalized over the amine, pyrene and partially also linker moieties, whereas the electron orbital remains localized on the pyrene core. This results in partial charge transfer from the linker region to the pyrene unit upon excitation. Consequently, the excitation energy is sensitive to the electronic properties of the linker. Both theoretical and experimental data confirm this trend. The computed absorption energy of the Ada-linked pyrene is lower by 0.103~eV relative to the tBuPh-linked analogue, consistent with the experimentally observed difference of approximately 0.075~eV. 

The higher-energy $S_0 \rightarrow S_2$ transition (3.5–4.0~eV) exhibits clear vibronic structure and corresponds to the $^1L_a$ excitation polarized along the longer molecular symmetry axis. Transition density and NTO analyses show that this excitation is localized entirely on the pyrene unit, with negligible linker influence, evidenced by only an 8~meV shift in the vibronic 0–0 transition. This is in agreement with the experimentally observed shift of 3~meV. Similar to the NDI and unsubstituted pyrene monomers, the computed reorganization energies for the C–C stretching modes are slightly overestimated, leading to stronger intensities of the higher vibrational transitions peaks compared to experiment. Nonetheless, the absolute excitation energies, spectral shapes, and fluorescence features are reproduced with excellent quantitative agreement. The lowest $S_1 \rightarrow S_0$ fluorescence transition, which retains partial ICT character, exhibits a slightly broader band than in experiment due to the same reorganization energy overestimation, but its energy and overall spectral profile are accurately reproduced. 

The excellent agreement between experimental and computed optical spectra for all monomeric systems validates the reliability of the employed computational protocol. This level of accuracy is particularly crucial for the subsequent modeling of hetero-PCPs, where differences in the absolute excitation energies of individual monomers directly determine the extent and character of electronic-state mixing.

\subsection{Optical properties of the homoparacyclophanes}

From the perspective of their optical and electronic properties, as well as for interpreting their spectra, the PCPs can be regarded as two optically active chromophoric units held in defined positions and orientations by two linkers. Within this framework, their excited-state properties can be investigated using two different approaches. 

The first is the super molecular approach, in which the entire PCP is treated as a single molecule, for which electronic and optical properties are computed directly. This approach provides the most accurate description of the system, as it naturally includes all mutual interactions between the two chromophoric units. However, this accuracy comes at a significant computational cost, particularly for the CC2 step, for which the scaling increases steeply with system size (roughly as N$^5$). The computational demand becomes even more severe when simulating transient absorption spectra, which require calculations of higher-lying singly and doubly excited states.

An alternative approach is the Frenkel exciton model, which treats the PCP as a dimer composed of two interacting monomeric units. Here, the aggregate Hamiltonian is expressed in a basis of separate non-interacting units. The diagonal elements correspond to the excitation energies of each unit within its environment, including the solvent and the remainder of the PCP molecule. The off-diagonal elements describe the Coulomb coupling between the transition densities of the two units, evaluated within their respective solvent cavities  (see Supporting Information for details). The limitations of this method arise when the inter-unit distance becomes small or when the excited states exhibit strong delocalization over the linkers. In such cases, a clear partitioning of the PCP into distinct, non-overlapping units becomes ambiguous, and a specific partitioning may be required for each electronic state. 

\begin{figure} [H] %[ht!]
    \centering
    \includegraphics[width=1.0\columnwidth]{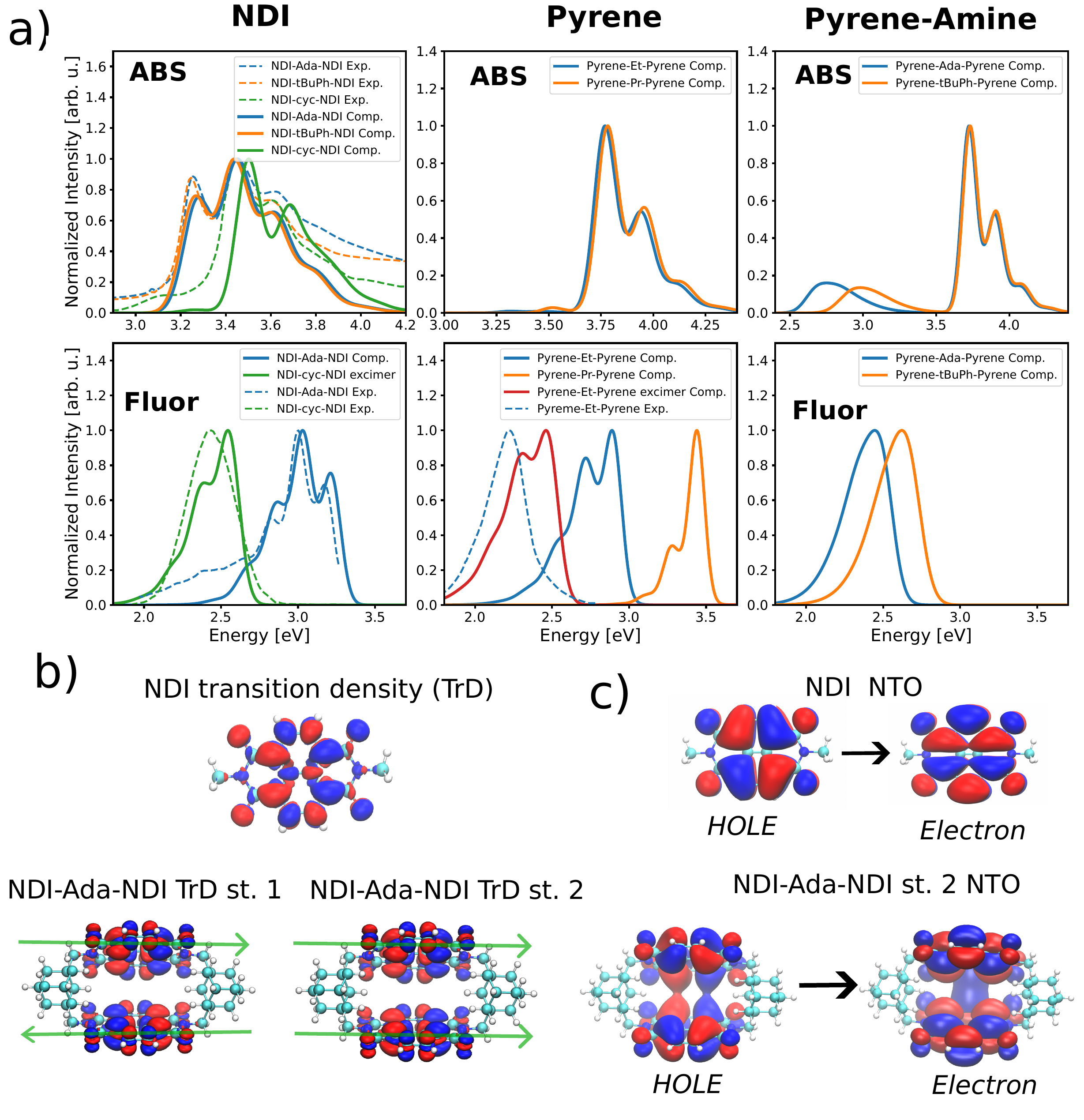}
    \caption{\label{fig:HomodimerSpectra} 
    \textbf{Optical spectra and corresponding electronic excitations of the homoparacyclophanes.} \textbf{(a)} Computed and experimental absorption and fluorescence spectra of NDI, pyrene and bis(aminomethyl)pyrene homo-paracyclophanes. \textbf{(b)} NDI and NDI-Ada-NDI PCP transition density for the corresponding lowest excited states. The green arrows represents the orientation of the transition dipole from the separate aromatic units. The lowest excited state of the H-type dimer is optically dark with unit transition dipoles oriented in opposite directions whereas the second excited state is bright with unit dipoles aligned. \textbf{(c)} Natural transition orbitals for the NDI and the second excited state for the NDI-Ada-NDI paracyclophane.
}
\end{figure}

In this section, we first demonstrate that the supermolecular approach reliably and quantitatively reproduces the experimental optical properties of the homo-PCPs. In fact, as for the case of the monomeric reference compounds, the computed absorption and fluorescence spectra agree very well with experimental measurements for both NDI- and pyrene- based homo-PCPs (Figure \ref{fig:HomodimerSpectra}). Even the vibronic structures are faithfully reproduced, with well-resolved progressions that closely mirror experimental spectra, where individual electronic and vibronic bands are often difficult to distinguish.

The optical spectra of the homo-PCPs can be interpreted in terms of excitonic H-type dimers of the two chromophoric units. Parallel stacking of the $\pi$-systems leads to a splitting of the lowest $\pi$–$\pi^*$ excitations in the dimers: a lower-energy, optically dark $S_1$ state and a higher-energy, bright $S_2$ state. Deviations from perfect parallel alignment arising from steric effects or geometric constraints of the linker lead to redistributions of the transition dipole strength between these two excitonic states, rendering the lower $S_1$ state partially allowed. This behavior is particularly evident in the NDI-based homo-PCPs. 

For the NDI–Ada–NDI and NDI–tBuPh–NDI systems, where the NDI units remain nearly parallel, the transition dipole moment of the $S_2$ state approaches the ideal H-type dimer limit, $\left|d^{PCP}_{02} \right|=\sqrt{2} \left|d^{NDI}_{01} \right| = 3.38 a.u.$, while $\left|d^{PCP}_{01} \right|\approx0$ for the $S_1$ state (a summary is given in Table S1 in SI). This assignment is also supported by the transition density analysis for NDI-Ada-NDI shown in Figure \ref{fig:HomodimerSpectra}b.
The dark $S_1$ state exhibits opposite transition dipole orientations on the two NDI units (antiparallel configuration), whereas the bright $S_2$ state shows parallel dipoles (constructive configuration). Small deviations from the idealized H-dimer behavior arise from weak molecular orbital (MO) overlap, linker effects, and differences in the local dielectric environment due to the presence of the other unit. 

In NDI–cyc–NDI, the two NDI units are rotated by approximately 9° with respect to each other, which prevents perfect cancellation of the transition dipoles in the $S_1$ "antiparallel" state, resulting in a weakly allowed transition. This is  manifested in the appearance of the small absorption peak near 3.1 eV corresponding to  the $S_{0}\rightarrow S_{1}$ transition, in addition to the dominant $S_0 \rightarrow S_2$ transition around 3.4 eV. Another important factor influencing the optical properties of PCPs is the inter-unit separation. NDI–Ada–NDI and NDI–tBuPh–NDI, with similar separations of 5.44 Å and 5.24 Å, respectively, show similar splitting energies and spectral shapes, whereas the shorter 2.9–4.0 Å separation in NDI–cyc–NDI increases the coupling strength, leading to larger exciton splitting, i.e. a blue-shift of the $S_2$ state and a red-shift of the $S_1$ state with respect to the Ada and tBuPh linkers.

The fluorescence spectra provide further insight into the nature of the lowest excited states. For NDI–tBuPh–NDI, the fluorescence intensity is significantly reduced relative to the monomer due to the weak transition dipole moment of the lowest exciton state, in line with the H-type dimer picture. The computed fluorescence spectrum within the vertical gradient approximation reproduces the experimental spectrum almost perfectly (Figure \ref{fig:HomodimerSpectra}), indicating minimal structural relaxation upon excitation. The fluorescence profile is thus essentially the mirror image of the absorption spectrum, shifted by the solvent reorganization energy. 

In contrast, the NDI–cyc–NDI system exhibits a large Stokes shift of approximately 0.65 eV between absorption and fluorescence maxima, which cannot be explained solely by solvent relaxation. The very short inter-unit distance, smaller than the sum of the van der Waals radii, introduces steric strain and mutual repulsion of the aromatic units in the ground state but favors excimer formation in the excited-state manifold. Excitation leads to charge redistribution and perfect paralell alignment with smaller separation of the NDI units in the optimal excited state geometry compared to the ground state one. Excitation relieves the ground-state strain due to the shorter optimal interunit distance, leading to the planarization of the aromatic units and a stabilization of the excimer configuration (see Figure~S2). The resulting structural relaxation significantly lowers the excited-state energy, causing the pronounced fluorescence red-shift. Theoretical simulations confirm the presence of an intra-PCP excimer in NDI–cyc–NDI, with the computed fluorescence spectrum for the relaxed structure in the excited state matching the experimental data (Figure \ref{fig:HomodimerSpectra}).

Excimer formation is not observed in NDI–tBuPh–NDI or NDI–Ada–NDI due to the bulky linkers maintaining interunit separations above 5 Å. At such distances, no ground-state strain/repulsion exists, and both ground and excited state geometries remain planar with just weakly interacting aromatic units. Excimer formation for these systems could occur only under high concentrations, where the interaction between aromatic units from different PCP molecules (rather than intra-PCP) become significant, due to short inter-PCP separation compared to the interunit separation within the single PCP molecule. 

Similar behavior is observed for pyrene-based PCPs. Systems with short ethylene linkers exhibit clear excimer fluorescence, whereas those with longer linkers show primarily monomer-like emission. In Figure \ref{fig:HomodimerSpectra}, the fluorescence spectra computed within the vertical gradient approximation are compared to those obtained after full excited-state relaxation. The relaxed (excimer) spectra are significantly red-shifted by around 0.5 eV compared to the ones computed within the vertical gradient approximation and they well reproduce the experimentally observed emission band. The slight blue-shift of the calculated spectra relative to experiment is consistent with the systematic overestimation of excitation energies by the employed combined TD-DFT and CC2 levels of theory, as noted above for the pyrene monomer. 

\begin{figure} [H] %[ht!]
    \centering
    \includegraphics[width=1.0\columnwidth]{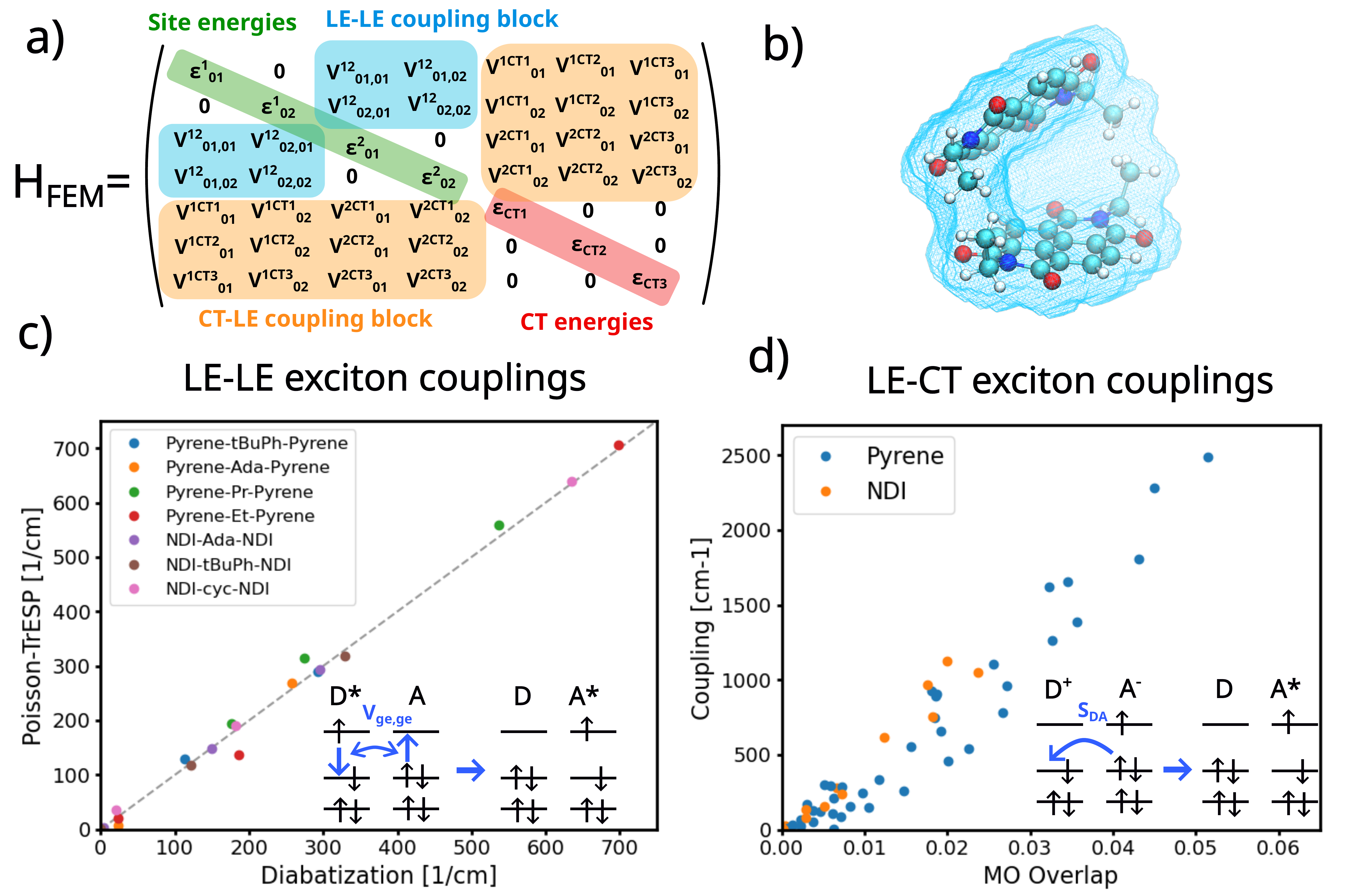}
    \caption{\label{fig:FrenkelExciton} 
    \textbf{Frenkel exciton model for the homo-paracyclophanes.} \textbf{(a)} Representation of the Frenkel exciton Hamiltonian and decomposition into LE-LE and CT-LE blocks. \textbf{(b)} Representation of the paracyclophane as two separate units embedded in a dielectric environment. The blue area represents the solvent cavity. \textbf{(c)} Comparison of the exciton couplings between locally excited states computed from the Poisson-TrESP method and from diabatization of the supermolecule excited states with correlation coefficient 0.997. \textbf{(d)} Correlation between calculated LE-CT couplings and corresponding molecular orbital overlaps between the individual separate units with correlation coefficient of 0.983 and 0.964 for the NDI and Pyrene based PCPs, respectively. 
}
\end{figure}

Having validated the supermolecular approach, we next assess the performance of the Frenkel exciton model in reproducing the same optical and electronic properties from monomer-level data and their mutual interactions. The key parameters in this model are the site energies (monomeric unit excitation energies) and the electronic couplings between the units. The electronic couplings can be divided into two categories: (i) interactions between locally excited states (LE–LE coupling) and (ii) couplings between locally excited and charge-transfer states (LE–CT coupling). The former can be expressed as Coulomb interaction of the corresponding transition densities, whereas the later one is related to molecular orbital overlap between the two units \cite{Sen2023}. The different form of coupling originates from the different physical nature of the processes. The LE-LE coupling can be understood as an excitation transfer process between the two units ($DA^{\ast}\rightarrow D^{\ast}A$), where the deexcitation of the molecule $A$ is accompanied by excitation of the molecule $D$. Such coupling originates from Coulomb interaction between transition densities and is often approximated within multipole expansion by dipole-dipole interactions (Förster resonant energy transfer). Here we adopted a more rigorous approach and computed electronic couplings via the Poisson-TrESP method \cite{ADOLPHS2006,Adolphs2008}, where transition densities are represented as distributed atomic transition charges in molecular cavities (Figure \ref{fig:FrenkelExciton}b), and their electrostatic interactions are evaluated in a dielectric environment using the APBS package \cite{APBS2018}. More details concerning the calculation of the electronic couplings can be found in the SI. 

The LE–CT coupling, corresponding to the process $D^{+}A^{-} \rightarrow DA^{\ast}$  during which the electron is transferred between the two units, depends linearly on the HOMO orbital overlap between donor and acceptor units, reflecting its one-electron transfer character. For the simplest case of the $HOMO^A \rightarrow LUMO^A$ excitation of the $A^{\ast}$ and  $HOMO^D \rightarrow LUMO^A$ excitation for the  $D^{+}A^{-}$ CT state, the process $D^{+}A^{-}\rightarrow DA^{\ast}$ corresponds to electron transfer from the $HOMO^A$ of the acceptor to the $HOMO^{D}$ of the donor. This electron transfer is therefore dependent on the overlap of $HOMO$ orbitals  of the donor and acceptor, yielding a linear dependence of the CT-LE coupling on the corresponding MO overlap. The graphical representation of the processes is shown in Figure \ref{fig:FrenkelExciton}. 

\begin{figure} [H] %[ht!]
    \centering
    \includegraphics[width=1.0\columnwidth]{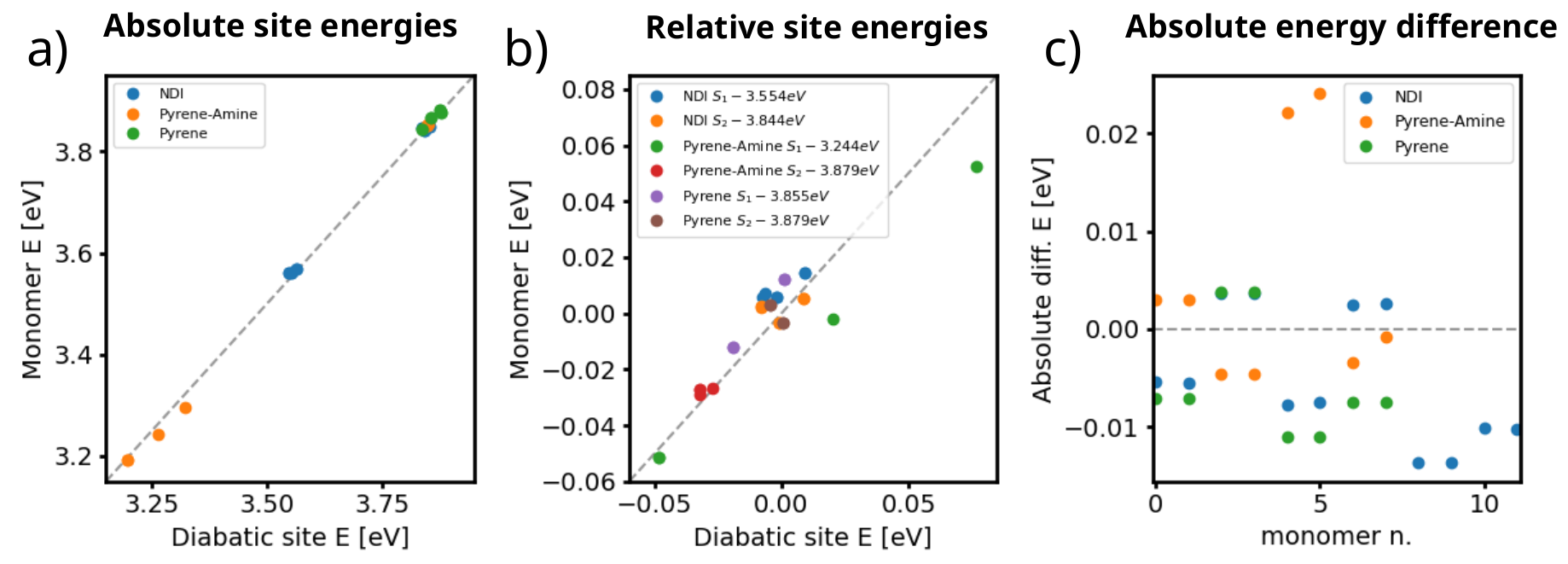}
    \caption{\label{fig:FrenkelExcitonE} 
    \textbf{Comparison of monomeric site excitation energies with the diabatic excitation energies.} \textbf{(a)} Absolute excitation energies from the aromatic monomeric units compared with the diabatic energies obtained from the FCD algorithm. \textbf{(b)} Absolute excitation energies shifted by mean diabatic excitation energy for each aromatic unit separately \textbf{(c)} Absolute energy difference between the diabatic states and monomeric excitation energies. 
}
\end{figure}

To compare the supermolecular approach with the Frenkel exciton model, the supermolecular excited states were diabatized using the Fragment Charge Difference (FCD) algorithm \cite{Nottoli2018}, yielding distinct LE and CT diabatic states and their mutual couplings. The resulting LE-LE couplings agree exceptionally well with those obtained from the Poisson-TrESP method across all the investigated PCPs, both for the NDI and pyrene systems (Figure \ref{fig:FrenkelExciton}). The CT–LE couplings also follow a nearly perfect linear dependence on the corresponding MO overlaps for both NDI and pyrene units with only small deviations. Remarkably, this agreement holds even for the shortest inter-unit distances such as in NDI–cyc–NDI and pyrene–Et–pyrene, where substantial orbital overlap is present.
The final parameter to be validated in the Frenkel exciton model is the site energy, corresponding to the excitation energy of each monomeric unit and forming the diagonal elements of the exciton Hamiltonian (Figure \ref{fig:FrenkelExciton}a). Site energies were computed for individual aromatic units within the PCP geometries (as used in the coupling calculations) using the same computational methodology as for the supermolecular approach. In this framework, each aromatic unit is embedded in a dielectric environment, and therefore mutual polarization effects between the two chromophores are neglected. Despite this simplification, the monomeric excitation energies agree exceptionally well with those obtained from diabatization of the supermolecular excited states  (Figure~\ref{fig:FrenkelExcitonE}). The absolute errors in excitation energies for the studied PCPs are typically within 10 meV, with one exception of pyrene–tBuPh–pyrene, which shows a deviation of 25 meV. This discrepancy likely arises from partial delocalization of the $S_{1}$ state through the linker and the truncation of the linker’s aromatic ring during separation into two monomeric units. Nevertheless, for the purpose of optical spectra simulations, such small deviations in site energies represent an excellent level of quantitative agreement.

Overall, these results demonstrate the robustness and reliability of the Frenkel exciton model in describing a wide range of homo-PCPs, from weakly coupled dimers to systems with partial charge-transfer character. It is also reasonable to expect that this approach will provide a quantitatively accurate description of hetero-PCPs in which the aromatic units exhibit small inter-unit molecular orbital overlap, as expected for inter-unit distances larger than the atomic van der Waals radii. Such conditions are fulfilled for example for PCPs with Ada and tBuPh linkers, however, new effects may arise for systems with shorter linkers, where significant strain due to the short distance between the aromatic units is present. This expectation is supported by our preliminary calculations, which will be discussed in detail in a forthcoming publication. Altogether, these findings validate the Frenkel exciton model as an efficient and physically transparent framework for understanding optical excitations and energy transfer processes in both homo- and hetero-paracyclophane architectures.

\subsection{Transient absorption}

\begin{figure} [ht!]
    \centering
    \includegraphics[width=1\linewidth]{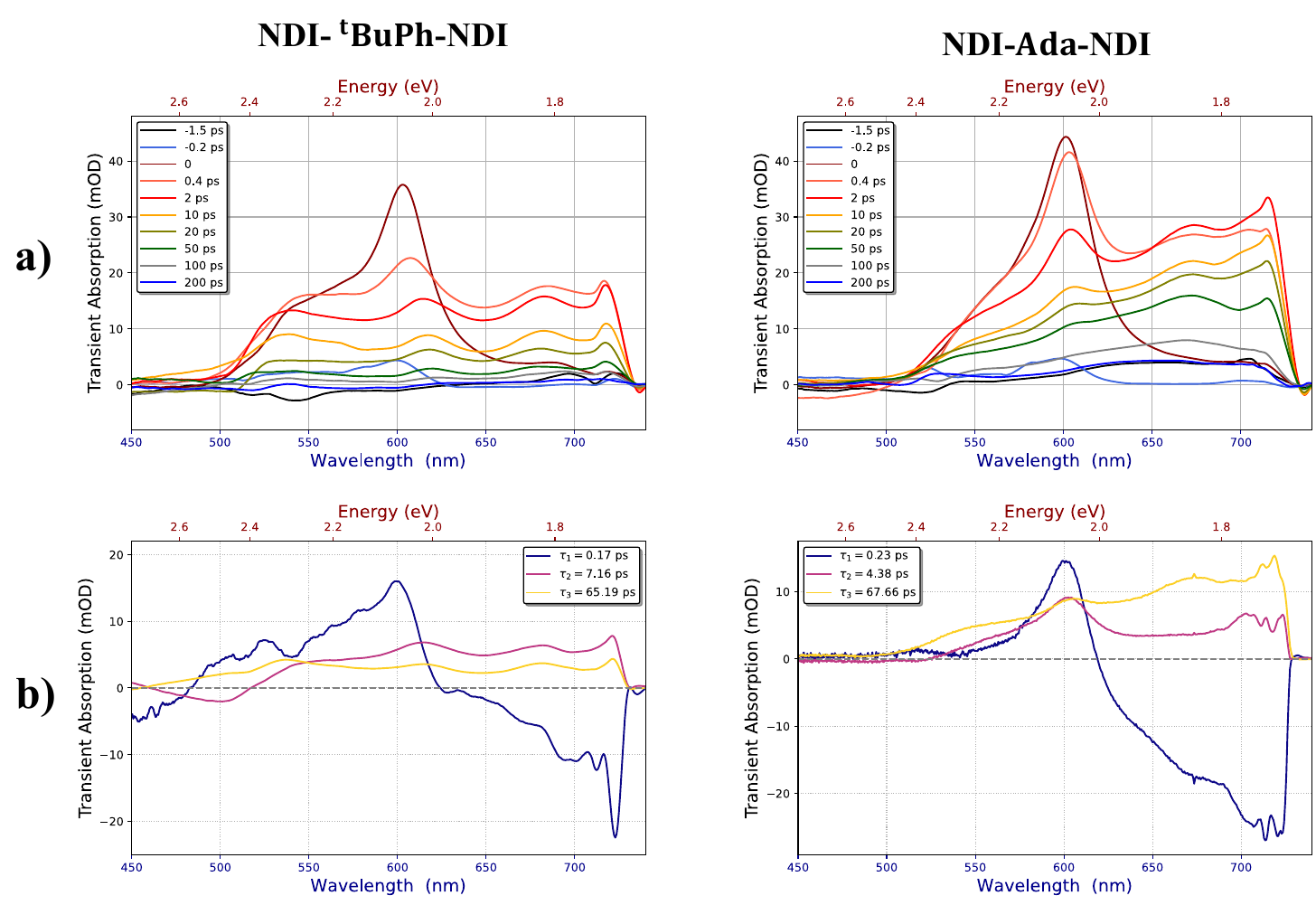}
\caption{\textbf{a)} Transient  absorption spectra of NDI-tBuPh-NDI and NDI-Ada-NDI recorded at various time delays excited at 380 nm and \textbf{b)} corresponding decay associated spectra (DASs).
}
\label{fig:Homodimer_spectra}
\end{figure}

To investigate the effect of the linker on the ultrafast excited-state dynamics of the NDI-based homodimers, femtosecond transient absorption (fTAS) measurements were performed with excitation at 380 nm, corresponding to the NDI $S_{1}$ absorption band. The transient absorption spectra recorded at different delay times for NDI-$^t$BuPh-NDI and NDI-Ada-NDI in \ce{CHCl3} are shown in Figure \ref{fig:Homodimer_spectra}\textbf{a}.

Immediately following photoexcitation, both compounds exhibit similar transient spectral features. The positive signal in the transient absorption spectra is attributed to excited-state absorption (ESA). At early delay times, the ESA displays a pronounced peak centered around 600 nm with a side band around 550 nm, whose relative intensity differs between the tBuPh- and Ada-linked PCPs. These features closely resemble the transient absorption spectrum of the NDI monomer \cite{Yushchenko2015} and can be assigned to the ESA from the NDI $S_{1}$ state.
At longer delay times, the ESA evolves into a much broader band with three prominent peaks around 550 nm, 610 nm and 680 nm. Such pronounced spectral changes cannot be explained solely by vibrational relaxation, but indicate substantial electronic relaxation. The broad and structured character of the spectra at longer delay times suggest the formation of a lower energy excimer state arising from interaction between the nearby NDI units of two separate PCPs. 

To quantify the underlying kinetics, the transient absorption data were analyzed using a global fitting procedure based on a sequential decay model. The best fit was obtained with three exponential decay components. The extracted lifetimes $(\tau)$ and their corresponding decay associated spectra are shown in Figure \ref{fig:Homodimer_spectra}\textbf{b}. For NDI-tBuPh-NDI, the time constants are $\tau_1=0.17$ ps, $\tau_2=7.16$ ps, and 
$\tau_3=65.19$ ps, while for the NDI-Ada-NDI corresponding vales are $\tau_1=0.23$ ps, $\tau_2=4.38$ ps, and $\tau_3=67.66$ ps.

 The ultrafast component $\tau_1$, represents the decay of the NDI $S_1$ ESA peak with a simultaneous rise of the broad band extending from 500 nm to 750 nm, near the edge of the spectral detection window. This process can be attributed to exciton relaxation from the lowest excited state $S_1$ of the H-type dimer within the PCP to the excimer state formed between adjacent PCPs. Such rapid excimer formation is facilitated by the high concentration of the sample (5 mM), which likely promotes the formation of $\pi$-stacked PCP aggregates already in the electronic ground state, before initial excitation. Ultrafast excimer formation has also been reported for the NDI-cyc-NDI PCP~\cite{Wu2015}, further supporting this assignment. 
 The intermediate component $\tau_2$ is associated with minor spectral evolution around 500 nm and may be attributed to vibrational relaxation in the excimer state. The longest-lived component $\tau_3$, can be interpreted as the decay of the excimer back to the ground state.

Although minor differences in the transient absorption spectra between Ada- and $^t$BuPh-linked PCPs are observed, their overall excited-state dynamics is remarkably similar. In contrast, previously reported transient absorption spectra of NDI-cyc-NDI in Ref.~\citenum{Wu2015}, indicate longer excimer lifetimes and slower relaxation to the ground state. This observation is consistent with the steady-state absorption and fluorescence data, as well as the computational results, which all suggest that the electronic and optical properties of the tBuPh- and Ada-linked PCPs are nearly identical, whereas the NDI–cyc–NDI system exhibits a substantially different behavior due to the shorter spatial separation between its chromophoric units. 

\subsection{Electrochemical properties}

The electrochemical properties of NDI-Ada-NDI, along with several monomeric reference compounds were investigated by cyclic voltammetry (CV), and differential pulse voltammetry (DPV) in \ce{CH2Cl2}  (Figures S9-S10). As summarized in Table \ref{tab:RedoxProp}, the NDI-Ada-NDI homodimer undergoes four reversible single-electron reduction processes at -0.62 V, -0.77 V, -1.21V and -1.25 V. It is expected that each NDI is sequentially reduced to the radical anion and the dianion species. The splitting of the first and second reductions suggests a strong electronic coupling between the two NDIs, resulting from the enforced interactions between their $\pi$-orbitals as previously reported for double-decker or multi-decker systems with NDI chromophores \cite{Keshri2020,Keshri2021}.

To gain deeper insight into the electrochemical behavior of the PCPs and the MOs involved in the redox processes, DFT calculations were performed. The geometries of the neutral, oxidized, and reduced species were independently optimized using the B3LYP functional with a 6-31+G(d,p) basis set including solvent effects from \ce{CH2Cl2} via the polarizable continuum model (PCM). The adiabatic ionization potentials (IP) and electron affinities (EA) were then derived from the optimized structures according to $IP=E^{opt}_{+1}-E^{opt}_{0}$ and $EA=E^{opt}_{0}-E^{opt}_{-1}$ where $E^{opt}_{i}$ denotes the total energy of the optimized structure with charge $i$. The computed IP and EA values show excellent quantitative agreement with experimental measurements, with deviations below 0.1 eV for all systems (Table \ref{tab:RedoxProp} and Table S2 in SI). Accurate reproduction of the experimental redox properties critically depends on the inclusion of solvent effects, particularly in polar environments. For charged species, solvent reorientation stabilizes the charge, resulting in molecular orbital energy shifts exceeding 1 eV. The HOMO and LUMO orbitals involved in the oxidation and reduction processes are shown in Figure S3 in the SI. Once again, the outstanding agreement between theory and experiment confirms the reliability of the computational methodology and its capability to accurately describe both geometric and electronic properties of PCPs.

\begin{table}[]
\resizebox{\columnwidth}{!}{%
\begin{tabular}{l|ccccc}
\hline
\multicolumn{1}{|l|}{\multirow{2}{*}{Compounds}} & \multicolumn{2}{c|}{Experimental}                                       &  \multicolumn{3}{c|}{Adiabatic PCM}                                                                             \\ \cline{2-6} 
\multicolumn{1}{|l|}{}                           & \multicolumn{1}{c|}{$-IP_{exp}$ (eV)} & \multicolumn{1}{c|}{$-EA_{exp}$ (eV)} & \multicolumn{1}{c|}{$-IP_{comp}$ (eV)} & \multicolumn{1}{c|}{$-EA_{comp}$ (eV)} & \multicolumn{1}{c|}{$E_{g,comp}$ (eV)} \\ \hline
NDI-Ada                                          &                                    &                                    & 6.81                               & 3.70                               & 3.11                               \\
NDI-tBuPh                                        &                                    & 3.82                               & 6.66                               & 3.80                               & 2.86                               \\
NDI-cyc                                          &                                    & 3.94$^\ast$                        & 7.04                               & 3.71                               & 3.33                               \\
Pyrene-Ada                                       & 4.76                               &                                    & 4.82                               & 1.94                               & 2.88                               \\
Pyrene-tBuPh                                     & 4.86                               &                                    & 4.98                               & 1.92                               & 3.06                               \\
Pyrene                                           & 5.58$^{\ast\ast}$                  &                                    & 5.56                               &  1.59                              &  3.62                              \\
NDI-Ada-NDI                                      &                                    & 3.79                               & 7.07                               & 3.81                               & 3.79                               \\
NDI-tBuPh-NDI-trans                              &                                    &  3.85                              & 6.73                               & 3.85                               & 2.88                               \\
NDI-cyc-NDI                                      &                                    &     4.27$^{\ast}$                  & 7.06                               & 3.93                               & 3.13                               \\
Pyrene-Ada-Pyrene                                &                                    &                                    & 4.76                               & 1.83                               & 2.93                               \\
Pyrene-tBuPh-Pyrene                              &                                    &                                    & 4.86                               & 1.81                               & 3.06                               \\
Pyrene-Et-Pyrene                                &                                    &                                    & 5.15                               &  2.30                        &  2.85                        \\
Pyrene-Pr-Pyrene                                 &                                    &                                    & 5.19                               & 2.07                               & 3.11                              
\end{tabular}
}
\caption{\label{tab:RedoxProp} \textbf{Redox properties for  homoparacyclophanes and corresponding monomeric reference compounds.} The experimental data were taken from $^{\ast}$ Ref. \citenum{Wu2014} and $^{\ast\ast}$ Ref. \citenum{Gac2025}.}
\end{table}

\subsection{Conclusions}

The excellent agreement between the simulated and experimental optical spectra demonstrates the robustness of the proposed computational methodology for a quantitative description of the optical and electronic properties of paracyclophanes (PCPs). Furthermore, representing PCPs as assemblies of interacting monomeric units has proven to be a powerful framework for interpreting both theoretical and experimental data. This fragment-based description also provides an efficient and accurate alternative for computing electronic and optical properties of a wide range of PCPs within the Frenkel exciton model.

Nevertheless, the fragment-based Frenkel exciton approach has inherent limitations. It cannot fully capture phenomena that intrinsically depend on dimer-specific characteristics, such as structural relaxation in the excited state leading to excimer formation. In such cases, the  computationally more demanding supermolecular approach remains necessary. Additional challenges may arise when combining CC2 and TD-DFT methods for systems where charge-transfer (CT) states are energetically close to and strongly mixed with the locally excited (LE) states. Since CC2 and TD-DFT can yield different relative energies for CT and LE states, their mixing and thus the resulting excited-state composition may differ. This issue can be mitigated by working in a diabatic basis, where pure CT and LE excitations are separated and the CC2 correction is applied prior to the back transformation to the adiabatic basis for the optical spectra calculations.

Despite these potential limitations, even for systems exhibiting mixed CT/LE character such as bis(aminomethyl)pyrene, the calculated spectra show excellent agreement with the experimental data. Therefore, we believe that cases where the proposed methodology fails are expected to be rather rare exceptions than common cases, confirming the broad applicability and reliability of this approach for the study of paracyclophane-based systems.

\begin{acknowledgement}

This work has been supported by the Swiss National Science Foundation grant numbers 200021-204053 and 200020-185092,  the SNSF Sinergia grant CRSII5-213533 as well as by the Werner Siemens Stiftung (WSS) for supporting the WSS Research Centre for Molecular Quantum Systems (molQ), and used computing time from the Swiss National Computing Centre CSCS.

\end{acknowledgement}

\begin{suppinfo}

The following is the supplementary material to this article:
Details concerning the calculation of the spectral densities, excitonic couplings effects of the Duschinsky rotation on the optical spectra and structures of the pyrene excimers. From the experimental methods, the synthetic procedures are discussed together with NMR data for the final compounds and cyclic voltametry measurements.

\end{suppinfo}

\bibliography{achemso-demo}

\end{document}


\section{Computational Methods}

\subsection{Spectral density}
The spectral density of the pigments was modeled by two contributions. The first contribution corresponds to the intramolecular vibrations coupled to the electronic excited states and is modeled by underdamped Brownian oscillator. The intramolecular normal modes were obtained with normal mode analysis and the corresponding Huang-Rhys factors through vertical gradient method \cite{Cignoni2022,Ferrer2012,Coker2016}. The structure of the system was optimized with the B3LYP functional with DCM solvent included through polarization continuum model (PCM). At the optimal geometry the normal mode analysis was performed using the same approach as for the geometry optimization. The gradients of the excited and ground state potential surfaces were computed at the TD-DFT approach with the range separated $\omega$B97XD functional which reasonably well describes both LE and CT states. The excited state gradients were corrected to the different DFT functional used for geometry optimization by subtracting non-zero gradient of the ground state. This is within the harmonic approximation used also for computing the spectral density exact. The second contribution originates from the nuclear motion of the environment and it is modeled by an overdamped Brownian oscillator. Reorganization energy of the environemnt was estimated to $\lambda \approx 200 cm^-1$ from PCM calculation with equilibrium and non-equilibrium solvation. The computed reorganization energy well represents the broadening of the vibronic peaks as well as the fluorescence stokes shift of the monomers. 

\subsection{Exciton couplings}
The exciton couplings between two LE states were computed within Poisson-TrESP approach. First the paracyclophane is separated into two aromatic units (A and B) containing also some linker atoms which contribute to the excitation of interest. Then the transition densities are computed for each monomeric units separately within TD-DFT approach inlcluding the solvent effects within the PCM method. The transition densities are than integrated into point transition charges localized on nuclei coordinates. Assignemnent of the transition density contribution to the atomic center was performed based on the closest distance. The resulting transition charges are rescaled to obtain the correct quantum mechanical transition dipole as:

\begin{equation}
  Q_{0j,a} = q_{0j,a} * \frac{ \left | \vec{d_0i}^{(QM)} \right |}{\left | \sum_b{q_{0j,b} \vec{r}_b} \right |},
  \label{eqn:AbsLine}
\end{equation}

where $q_{0j,a}$ correspond to transition charge for the excitation from ground to $j$-the excited state and located on atom $a$ at possition $\vec{r}_b$, the $\vec{d_0i}^{(QM)}$ represents the full qunantum chemical transition dipole and $Q_{0j,a}$ represents rescaled transition charges. The interaction energy in the dielectric environment was computed with APBS software package \cite{APBS2018}, were the rescaled transition charges are placed into vacuum cavities within dielectric environment. The cavities are formed by combined van def Waals surfaces of the two monomeric units. In the next step potential $\nu_{0j}^{(A)}  \left ( \vec{r}_b \right )$ of transition atomic charges from one unit ($A$) at the atomic positions $vec{r}_b$ of the other unit ($B$) was evaluated with optical dielectric constant of 2.032 for the dichloromethane environment. The final interaction energy is then computed as:

\begin{equation}
  V^{AB}_{0i,0j} = \sum_a Q^{(A)} _{0i,a} * \nu_{0j}^{(B)}  \left ( \vec{r}_a \right ),
  \label{eqn:AbsLine}
\end{equation}

\begin{figure} [H] %[ht!]
    \centering
    \includegraphics[width=1.0\columnwidth]{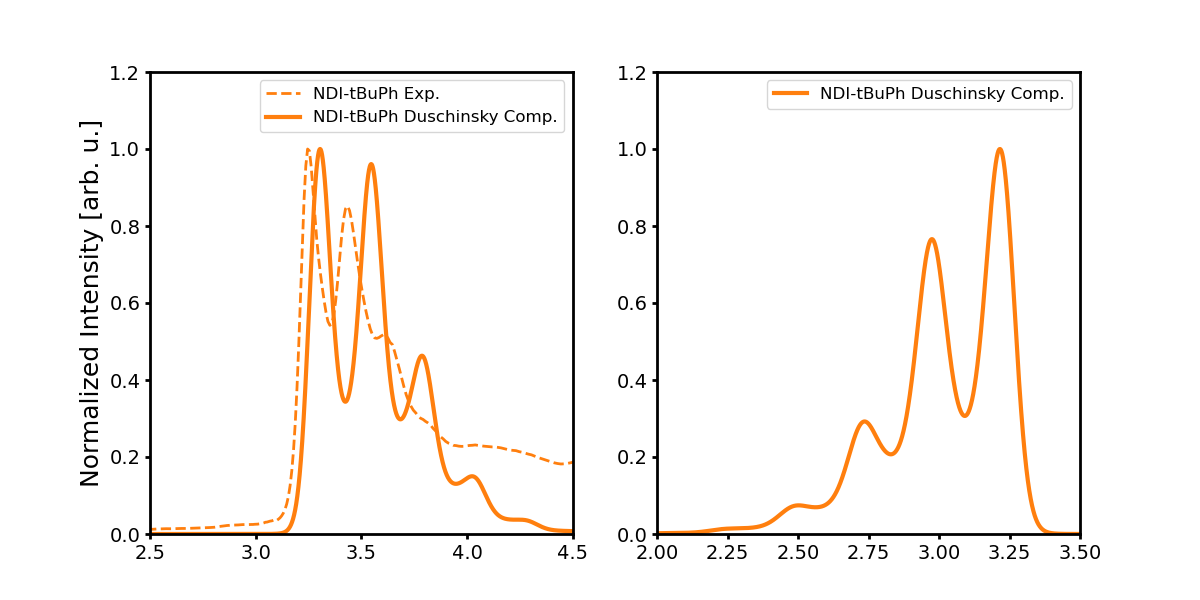}
    \caption{\label{fig:Duschinsky} 
    \textbf{Absorption and fluorescence spectra for the NDI computed with Duschinsky rotation.}  
}
\end{figure}

\begin{figure} [H] %[ht!]
    \centering
    \includegraphics[width=1.0\columnwidth]{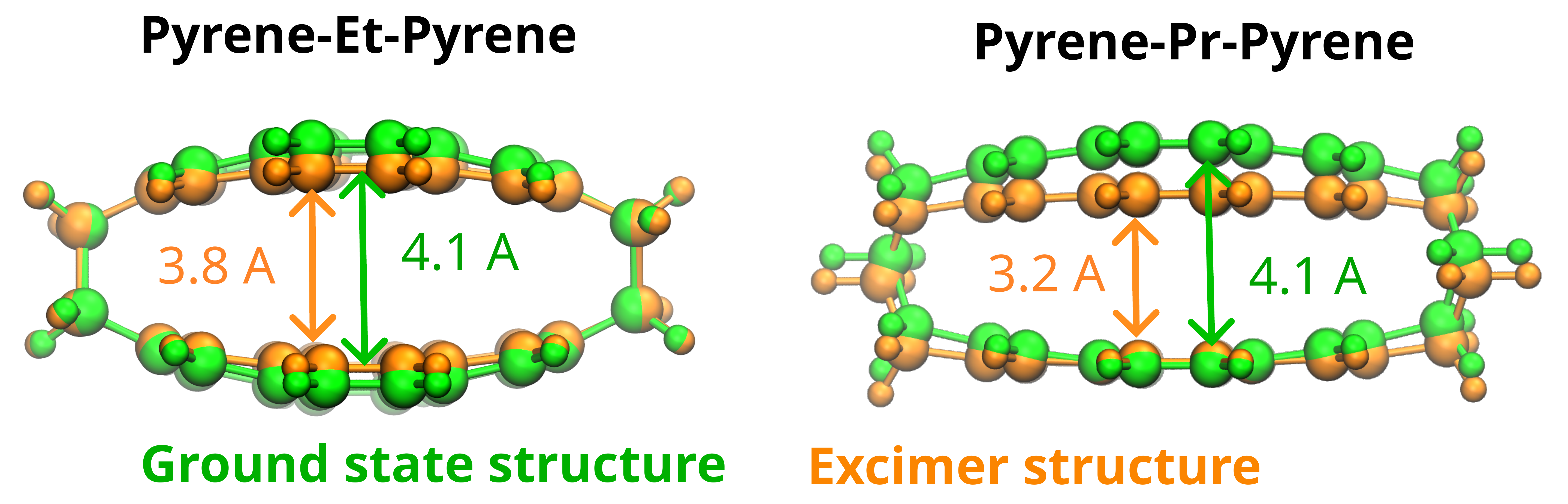}
    \caption{\label{fig:ExcimerGeom} 
    \textbf{Comparison of the ground state and excimer geometry for the pyrene homoparacyclophane with ethyl and propyl linkers.}  
}
\end{figure}

\begin{figure} [H] %[ht!]
    \centering
    \includegraphics[width=0.7\columnwidth]{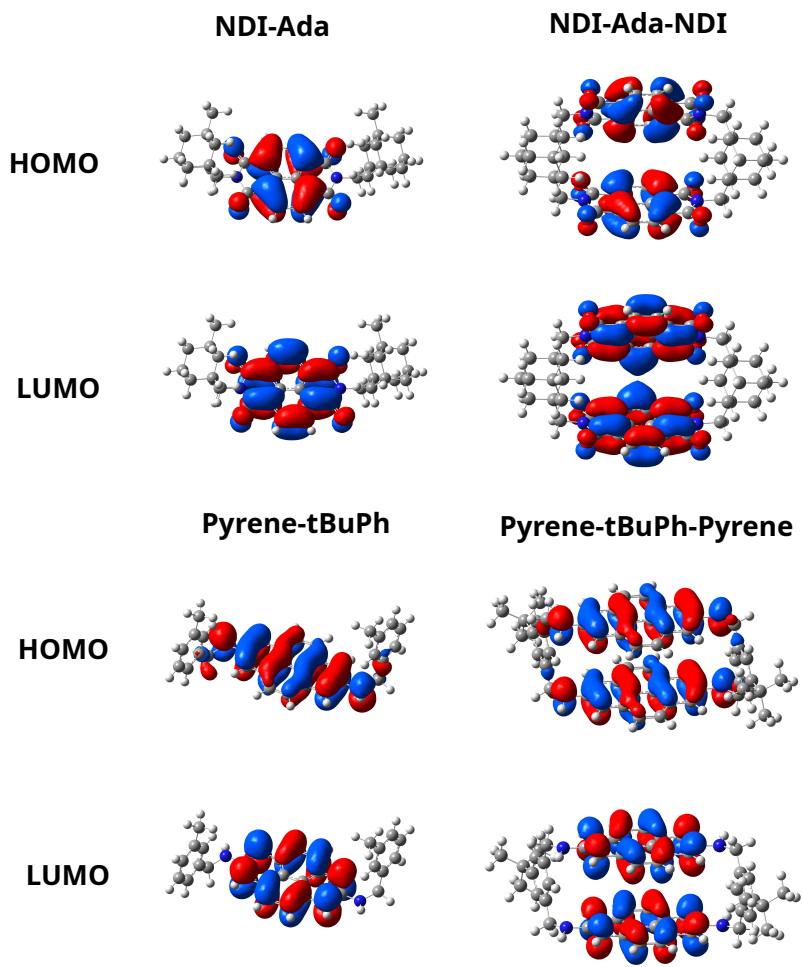}
    \caption{\label{fig:MOparacyclophanes} 
    \textbf{Representative examples of molecular orbitals for the NDI and pyrene reference compounds and homo-paracyclophanes.}  
}
\end{figure}

\begin{table}[]
\begin{tabular}{l|cccc}
Compound      & Energy $S_1$ {[}eV{]} & $\mu_{ge}$ $S_1$ {[}a.u.{]} & Energy $S_2$ {[}eV{]} & $\mu_{ge}$ $S_2$ {[}a.u.{]} \\ \hline \hline
NDI           & 3.205              & 2.39             &                    &                  \\
Pyrene-tBuPh  & 2.785              & 1.09             & 3.589              & 2.11             \\
Pyrene-Ada    & 2.682              & 1.10             & 3.586              & 2.06             \\
NDI-tBuPh-NDI & 3.168              & 0.0              & 3.252              & 3.47             \\
NDI-Ada-NDI   & 3.180              & 0.0              & 3.262              & 3.36            
\end{tabular}
\caption{\label{tab:Dipoles} Ab-initio adiabatic excitation energies and transition dipoles for reference monomeric compounds and paracyclophane homodimers.}
\end{table}

\begin{table}[]
\resizebox{\columnwidth}{!}{%
\begin{tabular}{l|cccccc}
\hline
\multicolumn{1}{|l|}{\multirow{2}{*}{Compounds}} & \multicolumn{3}{c|}{Vertical gas-phase}                                                                        & \multicolumn{3}{c|}{Vertical PCM}                                                                              \\ \cline{2-7} 
\multicolumn{1}{|l|}{}                           &  \multicolumn{1}{c|}{$-IP_{comp}$ (eV)} & \multicolumn{1}{c|}{$-EA_{comp}$ (eV)} & \multicolumn{1}{c|}{$E_{g,comp}$ (eV)} & \multicolumn{1}{c|}{$-IP_{comp}$ (eV)} & \multicolumn{1}{c|}{$-EA_{comp}$ (eV)} & \multicolumn{1}{c|}{$E_{g,comp}$ (eV)}  \\ \hline
NDI-Ada                                          & 8.0                                & 2.31                               & 5.68                               & 7.13                               & 3.68                               & 3.44                        \\
NDI-tBuPh                                        & 7.72                               & 2.45                               & 5.27                               & 6.81                               & 3.76                               & 3.06                        \\
NDI-cyc                                          & 8.68                                & 2.26                                & 6.40                               & 7.18                                & 3.56                                & 3.62                    \\
Pyrene-Ada                                       & 6.20                               & 0.31                               & 5.89                               & 5.05                               & 1.75                               & 3.31                        \\
Pyrene-tBuPh                                     & 6.34                               & 0.38                               & 5.96                               & 5.18                               & 1.79                               & 3.39                        \\
Pyrene                                           &  7.00                               & 0.29                                & 6.70                               & 5.67                                & 1.84                                & 3.38                    \\
NDI-Ada-NDI                                      & 8.53                                & 2.72                                & 5.62                               & 7.13                               & 3.68                               & 3.44                      \\
NDI-tBuPh-NDI-trans                              & 7.80                                    & 2.74                                    & 5.06                               & 6.81                               & 3.76                               & 3.06              \\
%NDI-tBuPh-NDI-cis                                    & 7.80                               &                                 & 6.23                                    & 2.25                                    & 3.98                               &                        &                                    &                                    & 4.97                               & 3.71                               & 1.24                               \\
NDI-cyc-NDI                                      & 8.45                                & 2.75                                & 5.70                               & 7.11                                & 3.82                                & 3.28                    \\
Pyrene-Ada-Pyrene                                & 5.76                                & 0.47                                & 5.30                               & 4.88                                & 1.79                                & 3.10                    \\
Pyrene-tBuPh-Pyrene                              & 5.86                                & 0.38                                &  5.48                              &  5.01                                    & 1.74                               & 3.27                \\
Pyrene-Et-Pyrene                                & 6.39                                & 0.74                                 & 5.65                               & 5.25                                & 2.08                                & 3.17                   \\
Pyrene-Pr-Pyrene                                 & 6.35                                & 0.66                                & 5.69                                & 5.28                                & 2.01                                & 3.26                     
\end{tabular}
}
\caption{\label{tab:RedoxPropSI} \textbf{Redox properties for  homoparacyclophanes and corresponding monomeric reference compounds.} The experimental data were taken from $^{\ast}$ Ref. \citenum{Wu2014} and $^{\ast\ast}$ Ref. \citenum{Gac2025}.}
\end{table}

\section{Experimental Section}

\subsection{Femtosecond Transient Absorption Spectroscopy (fTAS)}
Femtosecond pump-probe TAS was carried out with a Ti:sapphire regenerative amplifier with the central wavelength 800 nm, pulse duration 87 fs, repetition rate 4 kHz, and pulse energy 0.55 mJ.
\begin{figure} [ht!]
    \centering
    \includegraphics[width=0.7\linewidth]{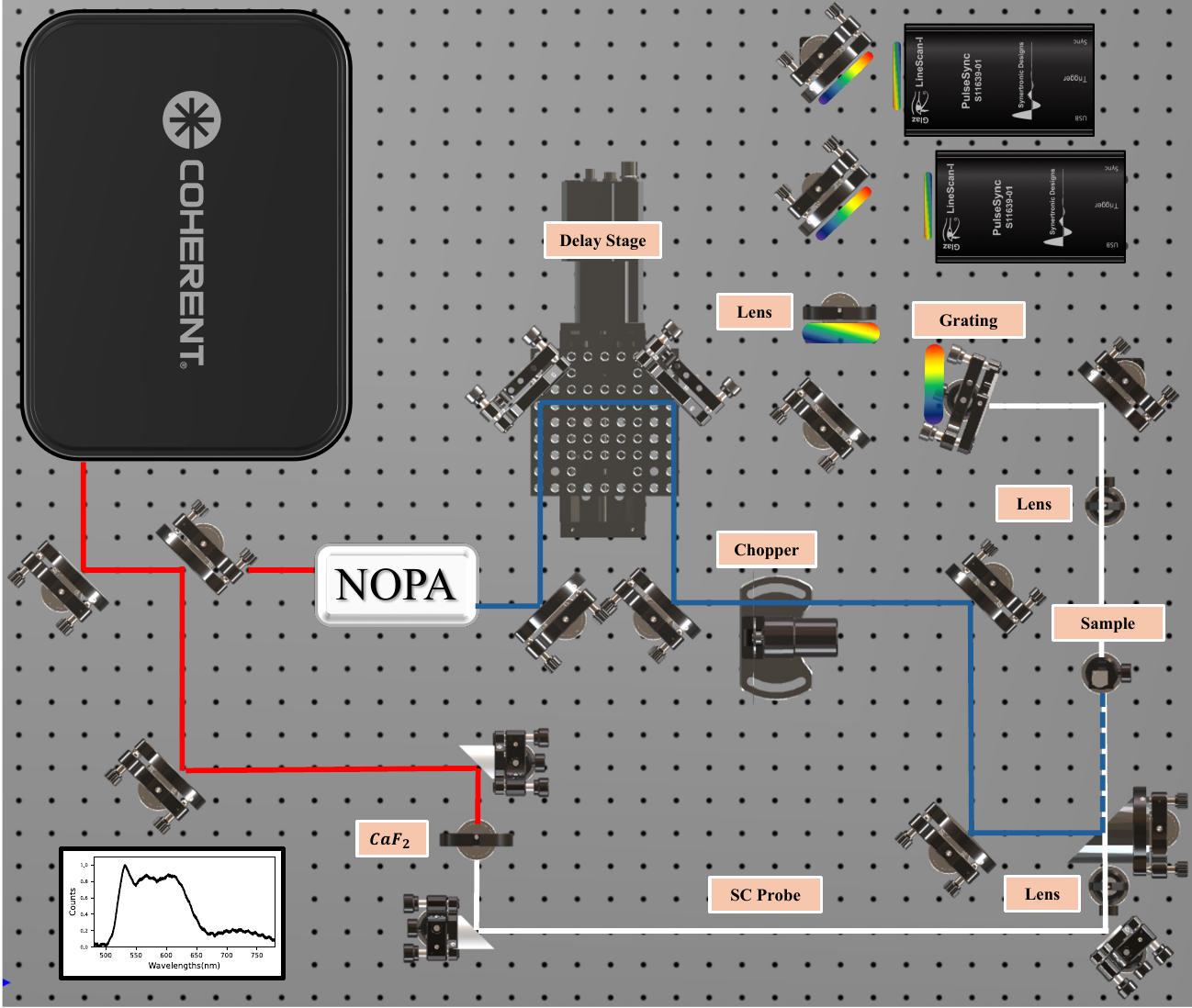}
    \caption{\label{fig:tas_setup} \textbf{Experimental setup for transient absorption spectroscopy (TAS) with a reference-corrected supercontinuum detection}. Pump pulses were generated in a noncollinear optical parametric amplifier (NOPA) and are spatially overlapped with femtosecond SC probe pulses generated in CaF$_2$. A mechanical chopper running at 2 kHz periodically interrupted the pump beam, allowing sequential acquisition of pump-on and pump-off traces for differential measurements.
}
\end{figure}

Figure \ref{fig:tas_setup} provides an overview of the experimental setup. Pump pulses were generated in a non-collinear optical parametric amplifier (NOPA) seeded with white light from the fundamental. The NOPA output at 760 nm was frequency-doubled in a 0.1 mm BBO crystal to produce excitation pulses centered at 380 nm.
The pump energy at the sample was adjusted to 0.5 $\mathrm{\mu}$J per pulse to avoid multiphoton artifacts.

Probe pulses were obtained by focusing [X] µJ of the 800 nm fundamental into a 5 mm CaF$_2$ plate, yielding a stable supercontinuum spanning 420-750 nm. The continuum was collimated and split into probe and reference arms to improve signal to noise ratio.
A mechanical chopper at 2 kHz modulated the pump beam for alternating on/off acquisition.
The instrument response function (IRF) was determined to be 200 fs (FWHM).
 Temporal delay between pump and probe was controlled by a motorized delay stage, covering to 1.9 ns.
Transient absorption spectra were recorded and detected on a dual CMOS cameras, achieving a spectral resolution of around 3 nm and a noise floor of around 100 $\mathrm{\mu}$OD. Prior to analyzing the data, the raw signal must be corrected for the group velocity dispersion (GVD) of the probe pulse. This procedure is necessary to establish an accurate and consistent time-zero reference point for all wavelengths across the entire probe spectrum. Then data were analyzed using global fitting with IRF convolution to extract decay associated spectra (DAS) and characteristic lifetimes.

\subsection{Paracyclophane structural analysys and sample preparation}

Air- and/or water-sensitive reactions were conducted under nitrogen and dry, freshly distilled solvents were used. Chemicals used to synthesize the compounds were purchased from commercial suppliers (Sigma-Aldrich, TCI, Alfa Aesar, or Thermo Fischer) and used without further purification. UV-Vis absorption spectra were measured on a Jasco V-730 spectrophotometer using quartz cuvettes with an optical path of 1 cm. Fluorescence spectra were collected on a Jasco spectrophotometer FP-8300. Automated flash chromatography (Silica gel) was performed with a Biotage Isolera Four instrument using a gradient of solvents that were given by the Rf values from the TLC. \ce{^{1}H} and \ce{^{13}C} NMR spectra were recorded on a Bruker Avance 300 or 400 spectrometer (\ce{^{1}H} NMR = 300 MHz or 400 MHz; \ce{^{13}C} NMR = 75 MHz or 101 MHz) Chemical shifts are reported in parts per million (ppm) and are referenced to the residual solvent peak (\ce{CDCl3}, $\delta$ \ce{^{1}H} = 7.26 ppm, and \ce{DMSO-d_6}, $\delta$ \ce{^{1}H} = 2.50 ppm; \ce{CDCl3}, $\delta$ \ce{^{13}C} =77.16 ppm, and \ce{DMSO-d_6}, $\delta$ \ce{^{13}C} = 39.52  ppm). The following abbreviations were used; s (singlet), d (doublet), and m (multiplet). High-resolution mass spectra (HR-MS) were measured by the Analytical Research and Services (ARS) of the University of Bern, Switzerland on a Thermo Fisher LTQ Orbitrap XL using Nano Electrospray Ionization. 
Cyclic voltammetry (CV) was performed in a three-electrode cell with a Pt working electrode, a glassy carbon counter-electrode, and an Ag/AgCl reference electrode. The electrochemical experiments were conducted under an oxygen-free atmosphere in dichloromethane with \ce{TBAPF_6} (0.1 M) as the supporting electrolyte. 

{\bf NDI-Ada-NDI} was synthesized by a procedure reported in the Ref~\citenum{Tominaga2018}, and the {\bf NDI-tBuPh} reference compound was synthesized by a procedure reported in the Ref.~\citenum{Gabutti2008}.

\subsection{Synthetic procedures of paracyclophane}

\begin{figure} [H] %[ht!]
    \centering
    \includegraphics[width=1.0\columnwidth]{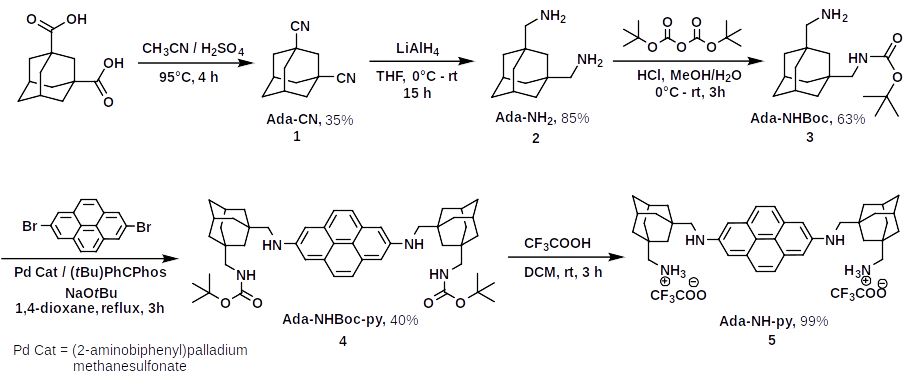}
    \caption{\label{fig:Synthesis} 
    \textbf{Synthetic route to homoparacyclophanes and monomeric units with linkers.}  
}
\end{figure}

\subsubsection{Synthesis of Ada-CN (1)}

n a 250 mL two-neck round-bottom flask equipped with a condenser, 1,3-Adamantanedicarboxylic acid (20 mmol, 4.48 g) was suspended in \ce{CH3CN} (120 mL), followed by dropwise addition of \ce{H2SO4} (22 mL, 98$\%$ wt/wt) at room temperature. The reaction mixture was refluxed under constant stirring for 4 h and then cooled to room temperature. The excess of \ce{CH3CN} was evaporated under vacuum, and the crude obtained was diluted with water (150 mL) and extracted with \ce{CH2Cl2} (150 mL). The aqueous phase was further extracted with two portions of \ce{CH2Cl2} (200 mL in total). The organic layers were combined and washed with saturated \ce{NaHCO3} followed by brine. The organic solution was dried over anhydrous \ce{Na2SO4}, and the solvent was evaporated to obtain a solid crude that was purified by automated flash column chromatography (Rf: 0.29, EtOAc: n-heptane 1:4). White powder, 1.35 g, 35$\%$ yield. \ce{^{1}H} NMR (300 MHz, CDCl3) $\delta$ 2.30 (s, 2H), 2.28 – 2.13 (m, 2H), 2.04 (d, J = 3.1 Hz, 8H), 1.73 (s, 2H).\ce{^{13}C} NMR (75 MHz, CDCl3) $\delta$ 123.17, 41.39, 38.49, 34.03, 29.89, 26.41. 

\subsubsection{Synthesis of Ada-\ce{NH2} (2)}

In a well-dried two-neck round-bottom flask, a suspension of \ce{LiAlH4} in THF (1.0 M, 25 mL) was added under a nitrogen atmosphere. The mixture was cooled to 0°C with an ice bath, followed by the slow addition of 1 (466 mg, 2.5 mmol) dissolved in dry THF (20 mL). After 30 minutes, the reaction was slowly warmed up to room temperature and left to react for a further 15 hours under constant stirring. The white suspension formed was exposed to air and then cooled down to 0°C. After that, it was carefully quenched with the following in this order: water (1 mL, added drop by drop), then NaOH (0.6 mL, 15$\%$ wt/v), and finally water (3 mL, vapors and gas are released during this process). The mixture was stirred for another 30 minutes, and the inorganic precipitate was filtered off using a glass frit (pore size 4) and washed with THF. The organic phase was dried over anhydrous \ce{Na2SO4} and then evaporated to obtain the product, which appeared as a colorless oil. No further purification was necessary. 0.44 g, 91$\%$ . Store the product under inert gas at a temperature below 4°C due to its instability under ambient conditions. \ce{^{1}H} NMR (300 MHz, \ce{CDCl_3}) $\delta$ 2.36 (s, 4H), 2.15 – 2.02 (m, 2H), 1.61 (m, 2H), 1.50 – 1.31 (m, 8H), 1.16 (m, 6H). \ce{^{13}C} NMR (75 MHz, \ce{CDCl_3}) $\delta$ 54.84, 42.57, 39.84, 36.96, 34.67, 28.70.

\subsubsection{Synthesis of Ada-NHBoc (3)}

In a two-neck round-bottom flask and under a nitrogen atmosphere, 1 equivalent of 2 (1.96 mmol, 0.38 g) was dissolved in \ce{MeOH} (15 mL). The solution was cooled to 0°C, followed by the slow addition of \ce{HCl} solution in diethyl ether (1.0 M, 1.96 mL). After that, the mixture was warmed to room temperature and stirred for 30 minutes. Subsequently, water (4 mL) was added and stirred for a further 30 minutes. Following this, a solution of 1.5 equivalents of \ce{(Boc)_2O} (2.94 mmol, 0.64 g) in \ce{MeOH} (20 mL) was added slowly over the course of 1 h, and the resulting solution was stirred for an additional hour. The mixture was concentrated under vacuum, then the residue was diluted with 10 mL of water. Next, the aqueous solution was extracted with two portions of \ce{Et_2O} (15 mL each) to remove unreacted amine and/or double-protected amine. After treating the aqueous phase with a solution of NaOH (2 equivalents in 10 mL of water), the monoprotected amine was extracted with \ce{CH2Cl2} (3 x 30 mL). The organic phase was washed with brine, dried over anhydrous \ce{Na2SO4}, and then evaporated to obtain a colorless oil, requiring no further purification. Oil, 0.40 g 69$\%$  yield. \ce{^{1}H} NMR (300 MHz, \ce{CDCl_3}) $\delta$ 4.57 (s, \ce{^{1}H}), 2.84 (d, J = 6.5 Hz, 2H), 2.38 – 2.36 (m, 2H), 2.08 – 2.06 (m, 2H), 1.68 – 1.59 (m, 5H), 1.46 – 1.35 (m, 16H), 1.20 – 1.17 (m, 2H). \ce{^{13}C} NMR (75 MHz, \ce{CDCl_3}) $\delta$ 156.48, 54.83, 54.69, 52.25, 42.56, 39.86, 39.83, 39.59, 36.95, 36.68, 34.67, 34.63, 28.69, 28.57, 28.52. HR-MS (ESI, positive): m/z calcd. for \ce{[C17H30N2O2+H]+} 295.2380; found: 295.2370.

\subsubsection{Synthesis of NDI-tBuPh-NDI}

A solution of 1,3-Bis(aminomethyl)-5-tert-butylbenzene (0.193 g, 1.0 mmol, 1 eq.) in warm anhydrous N, N-dimethylformamide (DMF, 2 mL) was slowly added to a solution of naphthalenetetracarboxylic dianhydride (267 mg, 1.0 mmol, 1 eq.) in anhydrous DMF (48 mL) at 130 °C under a nitrogen atmosphere. After 15 h, the reaction mixture was cooled down and subsequently concentrated under reduced pressure. The residue was purified by silica gel column chromatography using 10:1 DCM/EtOAc (Rf = 0.82) to give NDI paracyclophane as a pale-yellow solid (yield = 0.100 g, 23.6 $\%$). The product was recrystallized in a mixture of 1:1 chloroform and methanol (yield = 0.0441 g, 0.052 mmol, 10.4 $\%$). \ce{^{1}H} NMR (300 MHz, \ce{CDCl_3}) $\delta$ 8.60 (s, 8H), 7.44 (d, J = 1.6 Hz, 4H), 7.31 (t, J = 1.6 Hz, 2H), 5.34 (s, 8H), 1.35 (s, 18H). 

\begin{figure} [H] %[ht!]
    \centering
    \includegraphics[width=1.0\columnwidth]{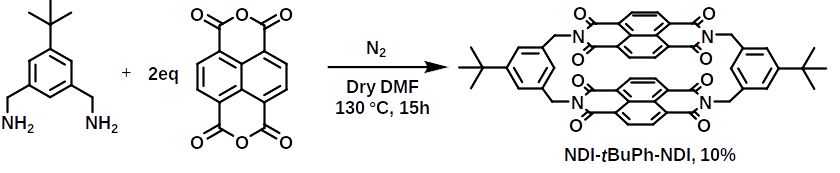}
    \caption{\label{fig:Synthesis_NDI-tBuPh-NDI} 
    \textbf{Synthetic route to NDI-tBuPh-NDI.}}
\end{figure}

\subsubsection{Synthesis of Pyrene-Ada}

A mixture of 2,7-dibromopyrene (240 mg, 0.68 mmol), NatBuO (156 mg, 1.63 mmol), 2-aminobiphenylpalladium methanesulfonate (10.5 mg), (tBu)PhCPhos (5.5 mg), and 3 (440 mg, 1.50 mmol) dissolved in dry 1,4-dioxane (10 mL) was stirred at 100°C under a positive flow of nitrogen gas. The mixture was monitored by TLC (EtOAc/n-heptane 3:7) and the reaction was stopped when the pyrene precursor was completely consumed. The reaction was cooled to room temperature and diluted with EtOAc. The solution was filtered through Celite and concentrated in vacuo. The crude was purified by automated flash column chromatography (Rf: 0.32, EtOAc: n-heptane 3:7). Yellow powder (220 mg, 41$\%$). \ce{^{1}H} NMR (300 MHz, \ce{CDCl_3}) $\delta$ 7.77 (s, 4H), 7.33 (s, 4H), 4.59 (s, 2H), 3.08 (s, 4H), 2.89 (d, J = 6.5 Hz, 4H), 2.13 (s, 4H), 1.69 – 1.60 (s, 14H), 1.49 – 1.40 (m, 30H). 
\ce{^{13}C} NMR (75 MHz, \ce{CDCl_3}) $\delta$ 156.48, 54.83, 54.69, 52.25, 42.56, 39.86, 39.83, 39.59, 36.95, 36.68, 34.67, 34.63, 28.69, 28.57, 28.52.
HR-MS (ESI, positive): m/z calcd. for \ce{[C50H67N4O4+]} = 787.5157; found: 787.5131. 

\begin{figure} [H] %[ht!]
    \centering
    \includegraphics[width=1.0\columnwidth]{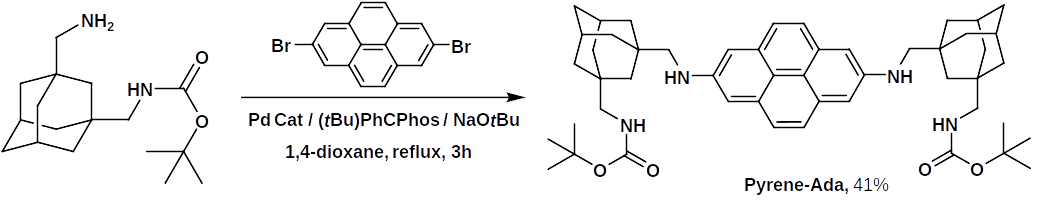}
    \caption{\label{fig:Synthesis_Pyrene-Ada} 
    \textbf{Synthetic route to Pyrene-Ada.}  
}
\end{figure}

\subsubsection{Synthesis of Pyrene-tBuPh}

\ce{Pd_2(dba)_3} (23.4 mg, 0.0256 mmol), brettPhos (27.5 mg, 0.0512 mmol), and NaOtBu (148 mg, 1.54 mmol) were suspended in 1,4-dioxane (5 mL) under a nitrogen atmosphere. Then, the suspension was heated at 65 °C for 90 minutes. After that, 2,7-dibromopyrene (230 mg, 0.64 mmol) was added, and subsequently, tBuPh-NHBoc (450 mg, 1.54 mmol) dissolved in 1,4-dioxane (5 mL). The reaction was stirred at 105 °C, monitored by TLC (EtOAc/n-heptane 3:7), and stopped when the pyrene precursor was completely consumed. The reaction was cooled to room temperature and diluted with EtOAc (30 mL). The solution was filtered through a pad of Celite and concentrated in vacuo. The crude was purified by automated flash column chromatography (Rf: 0.34, EtOAc: n-heptane 3:7). Yellow powder (145 mg, 29$\%$).\ce{^{1}H} NMR (400 MHz, CDCl3) $\delta$ 7.79 (s, 4H), 7.41 – 7.38 (m, 6H), 7.25 –7.21 (m, 4H), 4.82 (s, 2H), 4.53 (s, 4H), 4.33 – 4.31 (m, 4H), 1.44 (s, 18H), 1.32 (s, 18H). \ce{^{13}C} NMR (101 MHz, \ce{CDCl_3}) $\delta$ 156.07, 152.33, 145.77, 139.73, 139.27, 131.31, 127.20, 124.17, 123.75, 119.31, 110.06, 49.44, 45.08, 34.92, 31.54, 28.56. HR-MS (ESI, positive): m/z calcd. for \ce{[C50H63N4O4+]} = 783.4844; found: 783.4852.  

\begin{figure} [H] %[ht!]
    \centering
    \includegraphics[width=1.0\columnwidth]{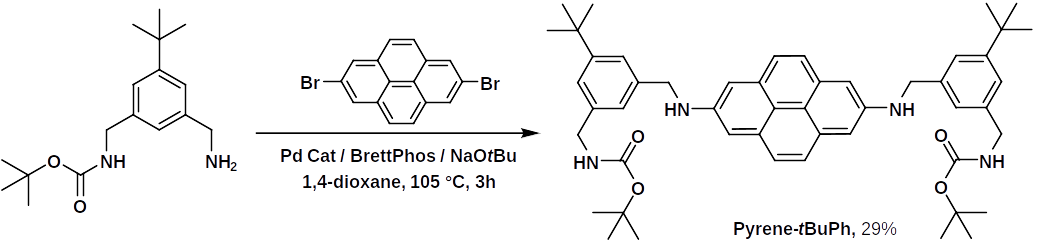}
    \caption{\label{fig:Synthesis_Pyrene-tBuPh} 
    \textbf{Synthetic route to Pyrene-tBuPh.}  
}
\end{figure}

\subsubsection{Cyclic voltammograms (CV) and differential pulse voltammograms (DPV) }

\begin{figure} [H] %[ht!]
    \centering
    \includegraphics[width=1.0\linewidth]{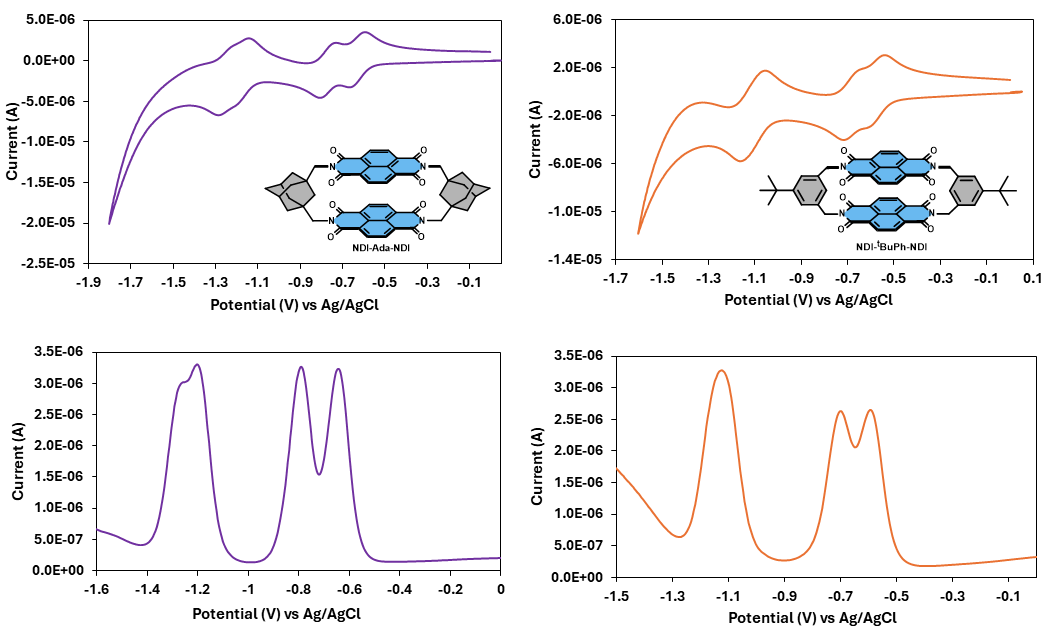}
    \caption{\label{fig:SI_CV_NDI-Ada-NDI} 
    \textbf{Cyclic voltammogram and differential pulse voltammogram of NDI-Ada-NDI (purple) and NDI-tBuPh-NDI (orange).}}
\end{figure}

\begin{figure} [H] %[ht!]
    \centering
    \includegraphics[width=1.0\linewidth]{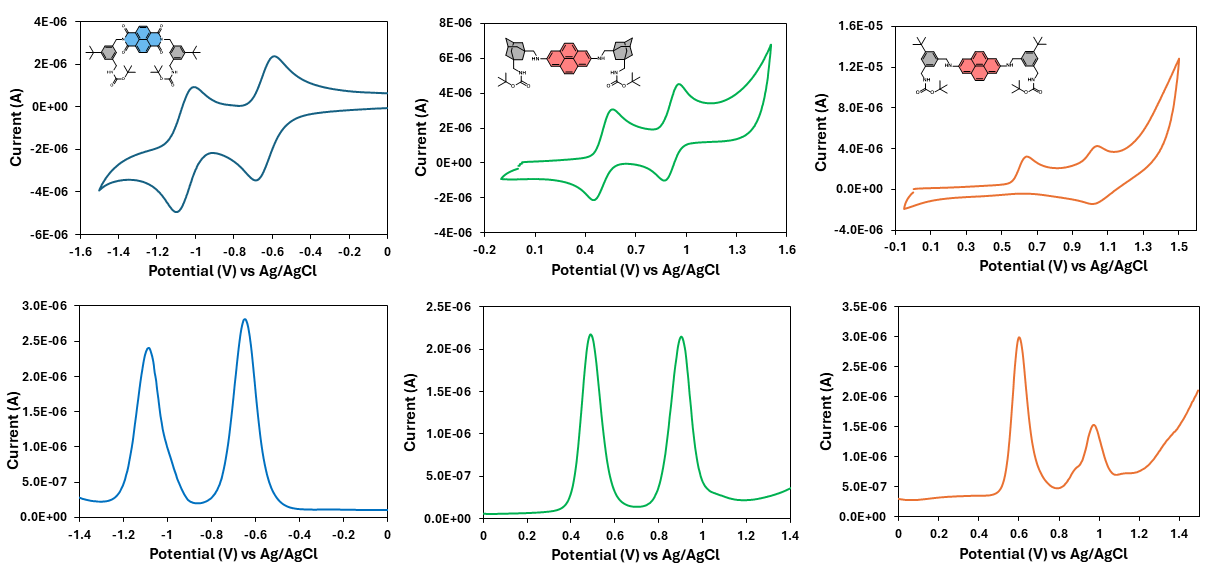}
    \caption{\label{fig:CV_monomers} 
    \textbf{Cyclic voltammogram and differential pulse voltammogram of NDI-tBuPh (blue), Pyrene-Ada (green) and Pyrene-tBuPh (orange).}  
}
\end{figure}

\begin{figure} [H] %[ht!]
    \centering
    \includegraphics[width=1.0\columnwidth]{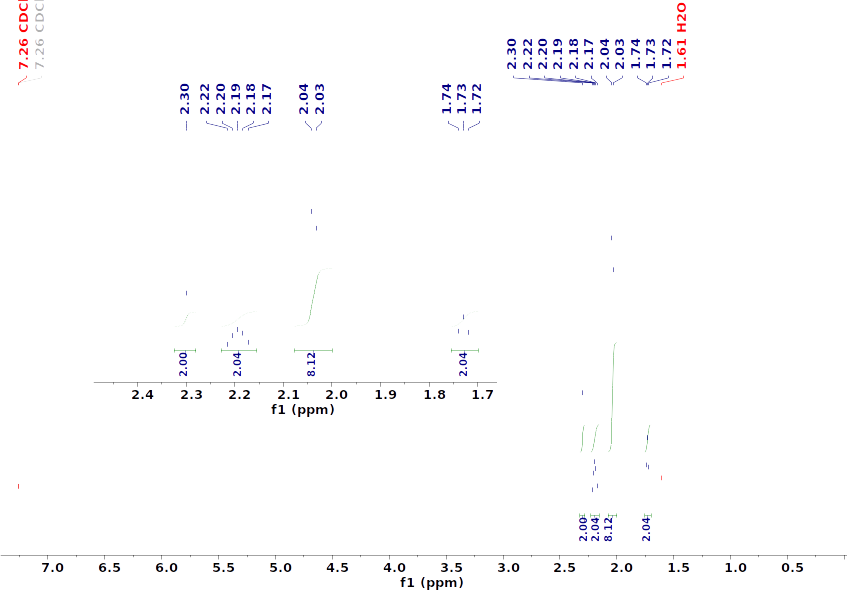}
    \caption{\label{fig:HNMR-Ada-CN} 
    \textbf{\ce{^1H}-NMR spectrum of Ada-CN.}  
}
\end{figure}

\begin{figure} [H] %[ht!]
    \centering
    \includegraphics[width=1.0\columnwidth]{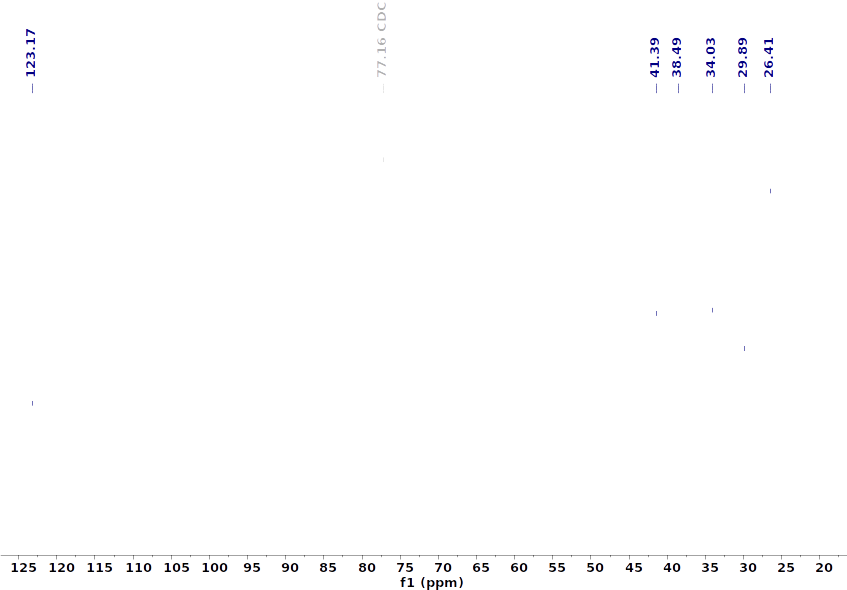}
    \caption{\label{fig:CNMR-Ada-CN} 
    \textbf{\ce{^{13}C}-NMR spectrum of Ada-CN.}  
}
\end{figure}

\begin{figure} [H] %[ht!]
    \centering
    \includegraphics[width=1.0\columnwidth]{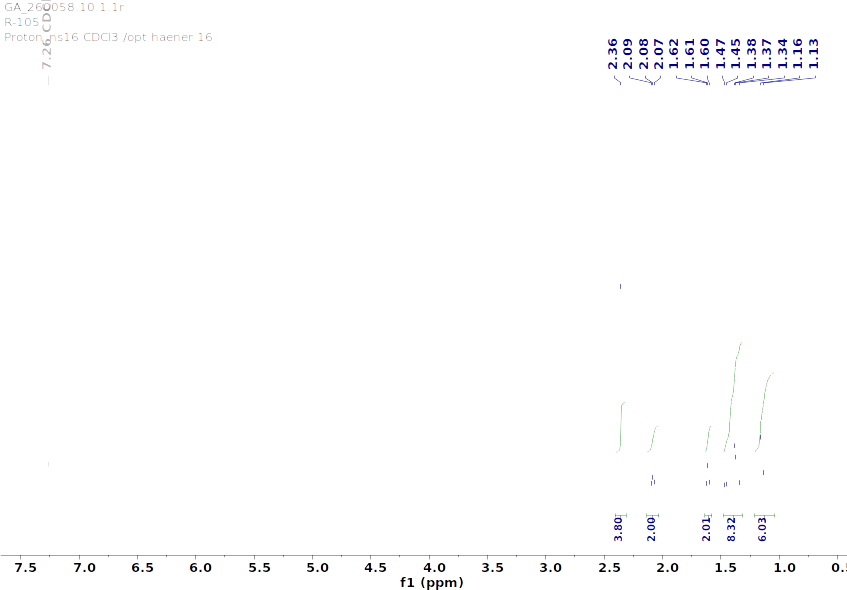}
    \caption{\label{fig:HNMR-Ada-NH2} 
    \textbf{\ce{^1H}-NMR spectrum of Ada-\ce{NH2}.}  
}
\end{figure}

\begin{figure} [H] %[ht!]
    \centering
    \includegraphics[width=1.0\columnwidth]{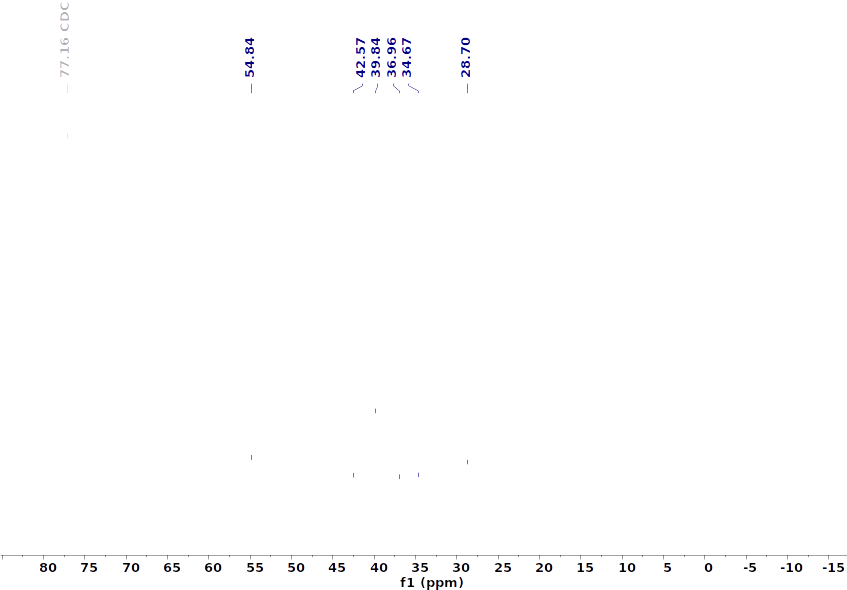}
    \caption{\label{fig:CNMR-Ada-NH2} 
    \textbf{\ce{^{13}C}-NMR spectrum of Ada-\ce{NH2}.}  
}
\end{figure}

\begin{figure} [H] %[ht!]
    \centering
    \includegraphics[width=1.0\columnwidth]{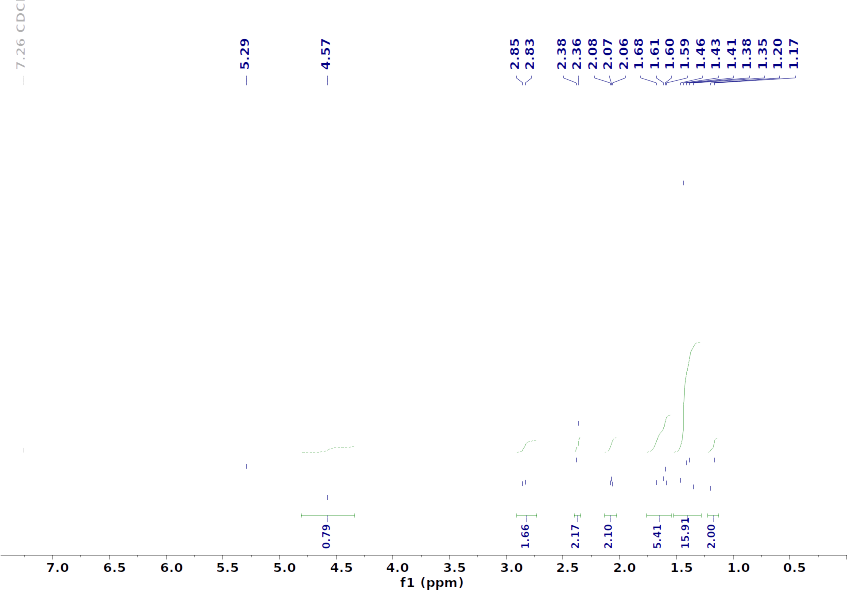}
    \caption{\label{fig:HNMR-Ada-NHBoc} 
    \textbf{\ce{^1H}-NMR spectrum of  Ada-NHBoc.}  
}
\end{figure}

\begin{figure} [H] %[ht!]
    \centering
    \includegraphics[width=1.0\columnwidth]{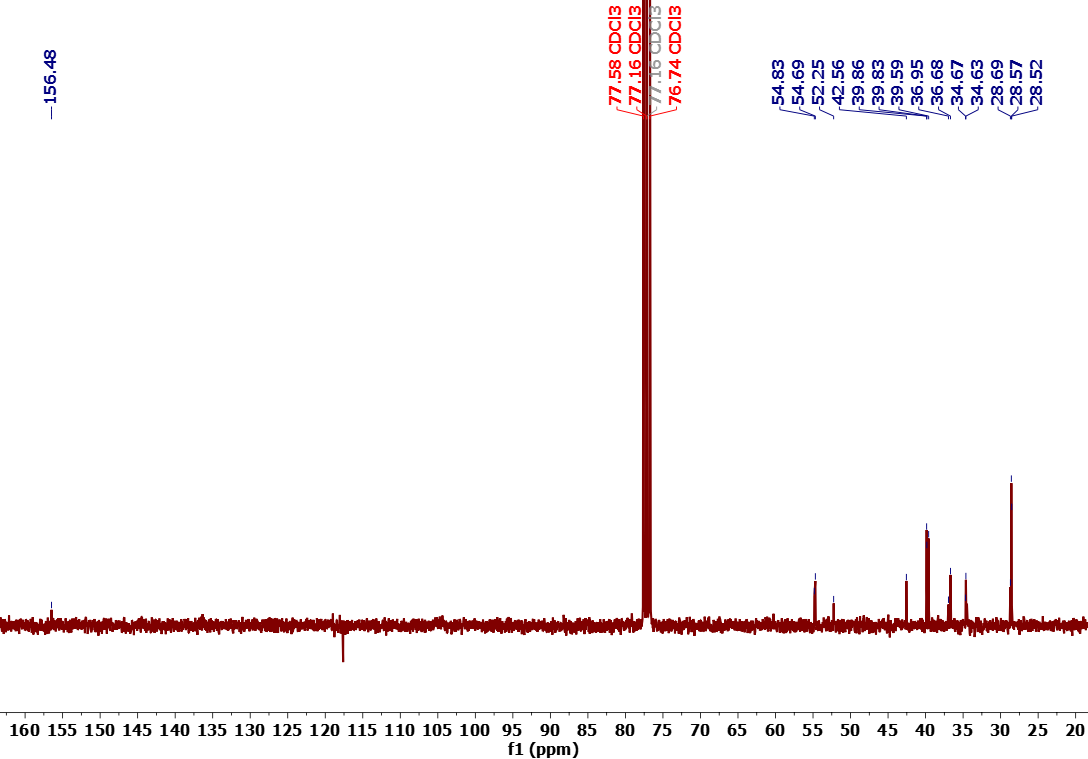}
    \caption{\label{fig:CNMR-Ada-NHBoc} 
    \textbf{\ce{^{13}C}-NMR spectrum of Ada-NHBoc.}  
}
\end{figure}

\begin{figure} [H] %[ht!]
    \centering
    \includegraphics[width=1.0\columnwidth]{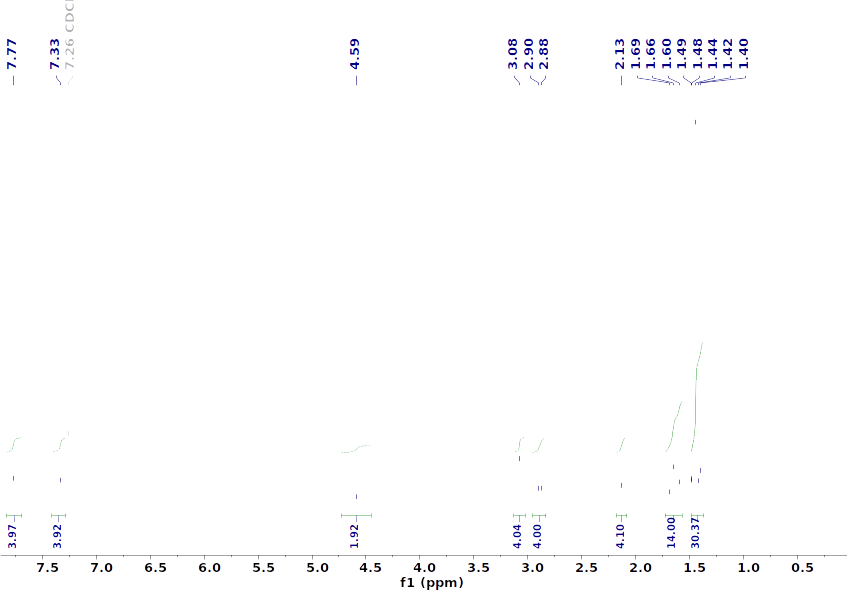}
    \caption{\label{fig:HNMR-Ada-NHBoc-py} 
    \textbf{\ce{^1H}-NMR spectrum of  Ada-NHBoc-py semicircle.}  
}
\end{figure}

\begin{figure} [H] %[ht!]
    \centering
    \includegraphics[width=1.0\columnwidth]{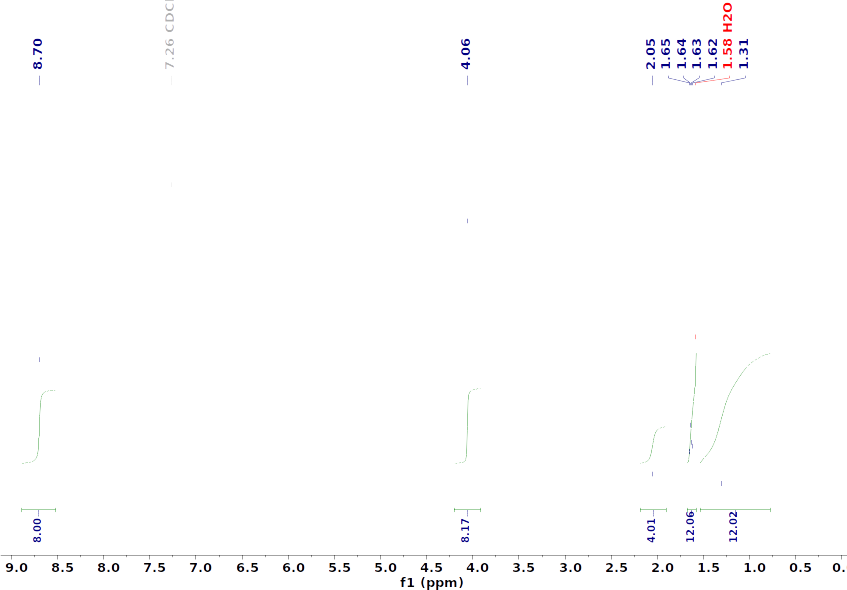}
    \caption{\label{fig:HNMR-NDI-Ada-NDI} 
    \textbf{\ce{^1H}-NMR spectrum of NDI-Ada-NDI.}  
}
\end{figure}

\begin{figure} [H] %[ht!]
    \centering
    \includegraphics[width=1.0\columnwidth]{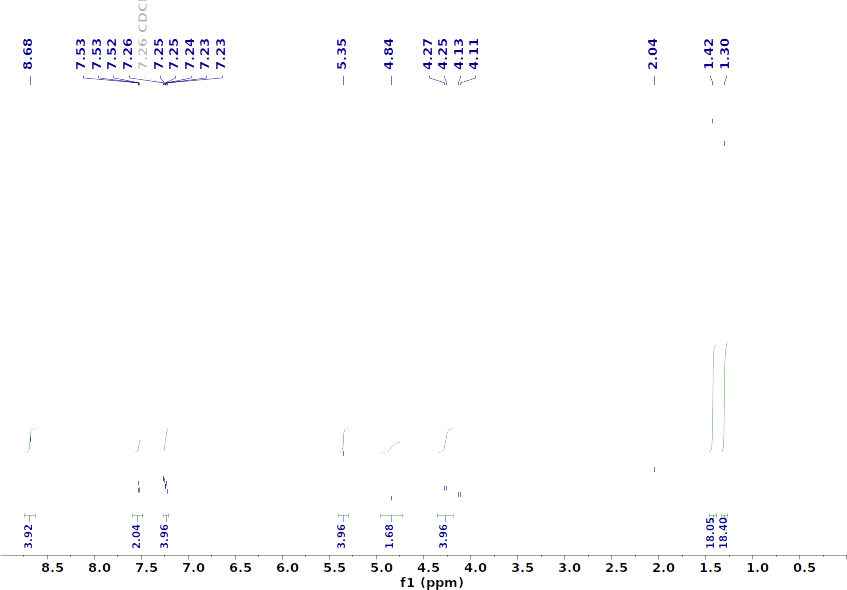}
    \caption{\label{fig:HNMR-NDI-tBuPh} 
    \textbf{\ce{^1H}-NMR spectrum of  NDI-tBuPh monomer.}  
}
\end{figure}

\begin{figure} [H] %[ht!]
    \centering
    \includegraphics[width=1.0\columnwidth]{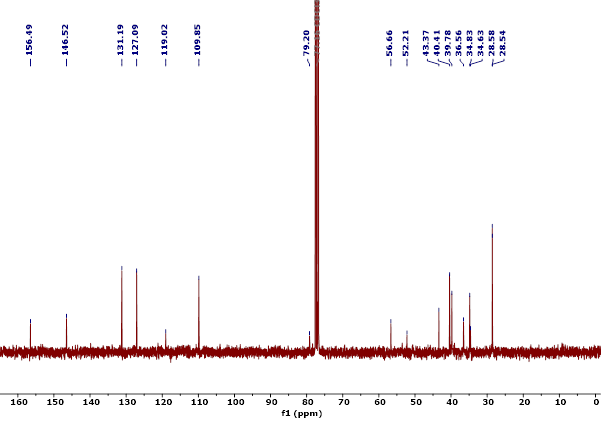}
    \caption{\label{fig:CNMR-Pyrene-Ada} 
    \textbf{\ce{^{13}C}-NMR spectrum of Pyrene-Ada monomer.}  
}
\end{figure}

\begin{figure} [H] %[ht!]
    \centering
    \includegraphics[width=1.0\columnwidth]{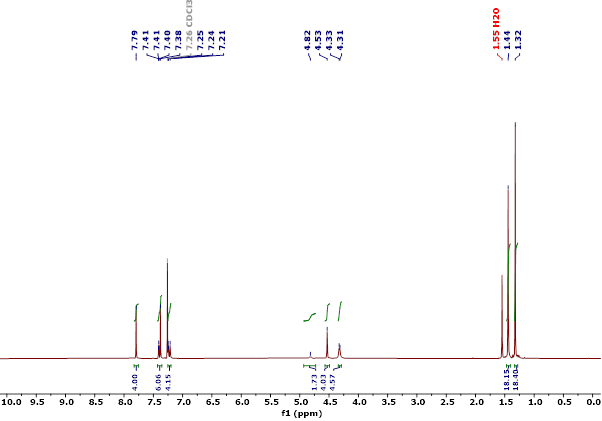}
    \caption{\label{fig:HNMR-Pyrene-tBuPh} 
    \textbf{\ce{^1H}-NMR spectrum of  Pyrene-tBuPh monomer.}  
}
\end{figure}

\begin{figure} [H] %[ht!]
    \centering
    \includegraphics[width=1.0\columnwidth]{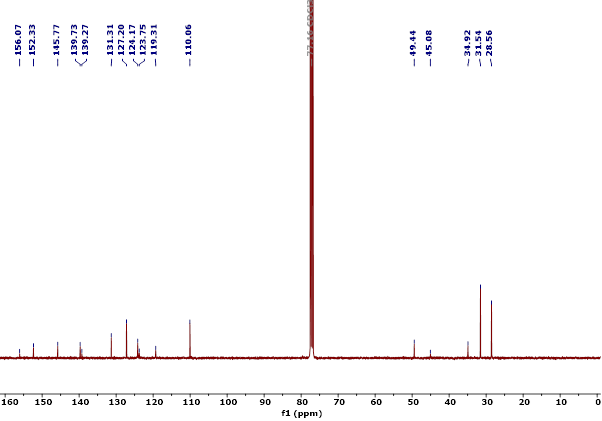}
    \caption{\label{fig:CNMR-Pyrene-tBuPh} 
    \textbf{\ce{^{13}C}-NMR spectrum of  Pyrene-tBuPh monomer.}  
}
\end{figure}

\bibliography{achemso-demo}